\documentclass[
 prx,reprint,
 floatfix,
superscriptaddress,
 amsmath,amssymb,
 aps,
 prx,
]{revtex4-2}

\usepackage{graphicx}
\usepackage{dcolumn}
\usepackage{bm}
\usepackage{svg}
\usepackage{microtype} 
\usepackage{paralist}
\usepackage[normalem]{ulem}
\usepackage{array}
\usepackage{siunitx}
\usepackage{wrapfig}
\usepackage{enumitem}
\usepackage{amsmath,amssymb}
\usepackage{array}
\usepackage{soul}
\usepackage{centernot}
\usepackage{bibunits}
\usepackage[titletoc, title]{appendix}
\usepackage{titletoc}
\usepackage{lineno}
\usepackage{lipsum}  
\usepackage{mathtools}
\usepackage{multirow}
\usepackage{graphicx}
\usepackage{upgreek}
\usepackage[unicode=true, colorlinks=true, citecolor={blue!80!black}, urlcolor={blue!50!black}, linkcolor = {blue!80!black}]{hyperref} 

\DeclareSIUnit\angstrom{\text {Å}}

\begin{document}

\title{
Resist-free shadow deposition using silicon trenches for Josephson junctions in superconducting qubits 
}

\author{Tathagata Banerjee}
\email{tb548@cornell.edu}
\affiliation{School of Applied and Engineering Physics, Cornell University, Ithaca, NY 14853, USA}

\author{Stephen Daniel Funni}
\affiliation{Department of Materials Science and Engineering, Cornell University, Ithaca, NY 14853, USA}

\author{Saswata Roy}
\affiliation{Department of Physics, Cornell University, Ithaca, NY 14853, USA}

\author{Judy J. Cha}
\affiliation{Department of Materials Science and Engineering, Cornell University, Ithaca, NY 14853, USA}

\author{Valla Fatemi}
\email{vf82@cornell.edu}
\affiliation{School of Applied and Engineering Physics, Cornell University, Ithaca, NY 14853, USA}

\begin{abstract}
    Superconducting qubit fabrication innovations continue to be explored to achieve higher performance. 
    Despite improvements to base layer fabrication and processing, resist-based Josephson junction (JJ) schemes have largely remained unchanged. 
    The polymer mask during deposition causes chemical contamination and limits \textit{in situ} and \textit{ex situ} surface preparation, junction materials, and scalability.
    Here, we demonstrate a resist-free approach to junction fabrication based on etched silicon trenches that is CMOS compatible and easily integrated into existing innovations in qubit base layer fabrication and chemical processing.
    We fabricate Al-AlO\textsubscript{x}-Al JJs and qubits using this method, measuring median energy relaxation times up to \SI{184}{\micro\second}.
    We find minimal contamination at the substrate-metal interface and fluctuations of energy relaxation on a 35 hour timescale that are narrow and normally distributed.
    The method widens the process window for substrate preparation and new materials platforms.
\end{abstract}

\maketitle

\section{Introduction}

Solid state qubit platforms offer scalable approaches to quantum computation.
Superconducting qubits in particular have demonstrated beyond break-even scaling of quantum error correction~\cite{sivak_realtime_2023,googlequantumaiandcollaborators_quantum_2025}. 
For the qubit array demonstrations of quantum error correction, a majority of the errors that limit performance of the error correction code arise due to the coherence times not being long enough compared to gate operation times~\cite{googlequantumaiandcollaborators_quantum_2025}. 
Therefore, it remains imperative to continue improving performance at the single qubit level. 

The last decade has seen substantial improvements to superconducting qubit coherence through innovations in materials choices and nanofabrication processing~\cite{bland_millisecond_2025,tuokkola_methods_2025, biznarova_mitigation_2024,bal_systematic_2024,gingras_improving_2025, olszewski_kryptonsputtered_2026}. 
In transmon qubits, these efforts have so far mainly targeted microwave losses originating at the surfaces of the `base layer' of the superconductor that account for the majority of the capacitance of the qubit. 
This has involved a multifaceted exploration on the choice of the base layer metallization, deposition methods, etching processes, and post-fabrication cleaning~\cite{bal_systematic_2024,gingras_improving_2025,olszewski_kryptonsputtered_2026,olszewski_lowloss_2025,banerjee_fabrication_2026, place_new_2021}. 
The investigations have so far culminated in transmon qubits with quality factors averaging \SI{15}{million} and energy relaxation times at the millisecond scale~\cite{bland_millisecond_2025,olszewski_kryptonsputtered_2026}, substantially surpassing the single-qubit coherence scale of scaled processors~\cite{googlequantumaiandcollaborators_quantum_2025}.
In these qubits, \textit{in operando}, the coherence times fluctuate over time by factors of 3 or more in a non-Gaussian distribution~\cite{berritta_realtime_2026,dane_performance_2025,abdisatarov_demonstrating_2025,gingras_improving_2025,tuokkola_methods_2025}. 
These fluctuations will limit the ability of superconducting qubits to scale up further as they will dominate errors and impose frequent recalibration duty cycles. 

\begin{figure*}
    \centering
    \includegraphics[width=\linewidth]{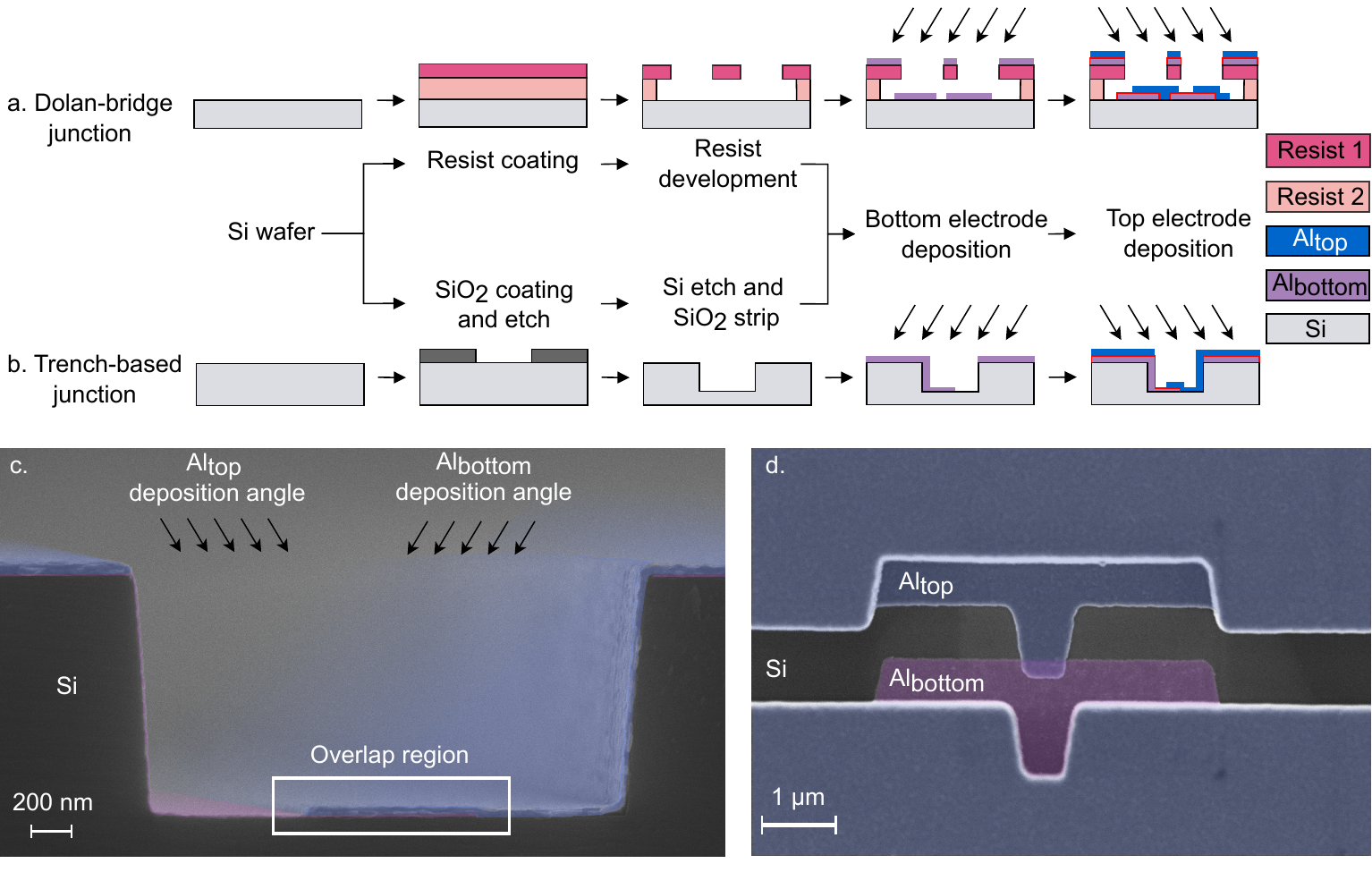}
    \caption{Josephson junction fabrication process. 
    (a) Dolan-bridge junction fabrication: a wafer is coated with double layer of photoresist, followed by lithography and development.
    This is followed by double angle deposition with an intermediate oxidation step to form the JJ.
    Finally, the junction undergoes a liftoff process to remove the resist (not shown).
    (b) Trench-based junction fabrication: a silicon wafer is thermally oxidized to form SiO\textsubscript{2}, followed by lithography and etching to form the hard mask.
    The Si is dry-etched followed by removal of the hard mask using hydrofluoric acid.
    This is followed by a double angle deposition with an intermediate oxidation step to form the overlap.
    (c) Cross-sectional false-colored electron micrograph of the JJ formed within the silicon trench.
    (d) Top-down false-colored electron micrograph of the JJ. 
    When the trench narrows, the shadowed region extends beyond the midpoint of the trench, leading to no overlap.
    Asymmetrical electrode shapes are designed to ensure minimal effect due to any rotational misalignment during loading and deposition.
    }
    \label{fig:JJfab_SEM}
\end{figure*}

In contrast to the base layers, the fabrication process for the Josephson junction (JJ) has remained largely unchanged over the last twenty-five years. 
Indeed, the materials used - aluminum metal electrodes and the native surface aluminum oxide for the tunnel barrier - have been identical since the first superconducting qubit~\cite{nakamura_coherent_1999}. 
This polymer liftoff-based fabrication process involves a double-angle deposition with an intermediate oxidation step, shown in Fig.~\ref{fig:JJfab_SEM}(a).
The principle advantage of this process is its simplicity, with a single lithography step and minimal number of process parameters, which makes process optimization relatively straightforward.
However, the presence of the polymer mask results in chemical contamination and severely limits options for surface preparation, junction materials, deposition methods, and other fabrication process windows.
Other fully subtractive methods, such as the trilayer process, have been explored~\cite{choi_low_2025,anferov_improved_2024,anferov_superconducting_2024,verjauw_path_2022}, 
but so far they are hindered by the substantially larger number of process steps which must be optimized to achieve sufficiently high quality factor devices to be competitive.

The limitations of both the liftoff and subtractive processes in use today promotes the opportunity for an approach that fulfills the need for broader process exploration and optimization for junction fabrication while maintaining the simplicity of single-step shadow-evaporation.
Ideally, such a process would be relatively straightforward, involve only a few nanofabrication steps, be compatible with now-standard angle evaporation tools, and take advantage of a fully inorganic, process resilient environment.
Efforts with inorganic stencil masks have recently aimed to accomplish this, but mechanical challenges and large mask-substrate distances limited fabrication precision and have not yet shown improvement~\cite{hanna_onchip_2026,tsioutsios_freestanding_2020}. 
So far, a simple, nanometrically precise approach that demonstrates leading-edge performance has remained elusive.

Here, we demonstrate a new, relatively simple approach to accomplishing single-step Josephson junction formation with only a trenched silicon substrate to enable the shadowing effect.
The preparation of the substrate trench uses standard, clean complementary metal-oxide-semiconductor (CMOS) methods. 
The JJ deposition can then occur after surface preparation to ensure full removal of the surface oxide.
We fabricate Al-AlO\textsubscript{x}-Al JJs with no detectable carbon or oxygen at the metal-substrate interface.
Finally, we conduct a subtractive-only process to define transmon qubits with quality factors as high as 3 million ($T_1$ averaging \SI{184}{\micro\second}).
Our novel, clean, simple nanofabrication approach opens a broad landscape of processing conditions, surface preparation, and materials choices.

\section{JJ Fabrication and measurement}

The junction fabrication scheme is based on etching trenches into the Si substrate. 
The sidewalls of the trench provide the necessary shadowed area for an overlap region to form the junction while leaving non-overlapped regions for the two electrodes to climb the sidewall onto the top of the substrate. 
The formation of the overlap is controlled by modulating the width of the trench.
When the trench narrows, the shadowed region extends beyond the midpoint of the trench, at which point no overlap is formed. 
Thus, we can define the width of the junction changing by the width of the trench, and the length of the junction by the length of the wider trench before it narrows.
 
The chip is first prepared by etching the trench into the substrate, as shown in Fig.~\ref{fig:JJfab_SEM}(b).
We thermally oxidize high-resistivity Si(100) to grow SiO\textsubscript{2} for a hard mask. 
We define an etch mask with an electron-beam lithography process, followed by a CHF\textsubscript{3}/O\textsubscript{2}-based dry-etch process to transfer the pattern onto the hard mask.
The Si is then etched with the SiO\textsubscript{2} hard-mask using an HBr/Ar etch designed to maintain a smooth silicon surface. 
Finally, the SiO\textsubscript{2} is stripped using 10:1 hydrofluoric acid (HF).
We note that many variants of silicon etch could be used to accomplish this step, with all steps being fully CMOS compatible.
See Appendix~\ref{app:trenchfab} for further details of the trench etching process.

Immediately prior to junction formation, the chip is soaked in 10:1 buffered oxide etch (BOE) to strip any native SiO\textsubscript{2} and remaining resist residue to ensure a clean substrate.
The \SI{20}{\nano\meter} (\SI{80}{\nano\meter}) aluminum for the bottom (top) electrode is evaporated at \SI{49}{\degree} (\SI{-49}{\degree}) at a rate of \SI{2}{\angstrom/\second}. 
Between depositions, the bottom electrode is oxidized at \SI{1}{Torr} for \SI{5}{\minute} using $100\%$ O\textsubscript{2} (Appendix~\ref{app:JJfab}).
Fig.~\ref{fig:JJfab_SEM}(c) shows a cross-sectional SEM image of a \SI{3}{\micro\meter} wide trench with the evaporated layers and a clear overlap region.

To form the JJ without any additional subtractive processing, the trench has a variable width. 
This is shown in Fig.~\ref{fig:JJfab_SEM}(d) with a top-down SEM image of a JJ formed by the process. 
The overlap region is formed at the widest part of the trench, and narrower parts of the trench result in only single layers of aluminum elsewhere in the trench or on the sidewalls.
The asymmetric electrode shapes are chosen to ensure that small rotational misalignment of the chip during deposition produces minor changes to the junction shapes.
We also remark that we do not observe the `halo' of hydrocarbons often seen when using resist-based liftoff~\cite{pop_fabrication_2012}.

Room temperature two-probe measurements are conducted on junctions of varying areas and oxidation doses (see Appendix~\ref{app:JJ}). 
The JJ shown in Fig.~\ref{fig:JJfab_SEM}(d), with dimensions of \SI{360}{\nano\meter}$\times$\SI{180}{\nano\meter} and the above dose, has a resistance of around \SI{6.6}{\kilo\ohm} and a calculated critical current density $J_C$ of $72.2\pm 6.2$~\si{\ampere\per\square\centi\meter}. 
This is comparable to values seen by others, and in the range for transmon qubits, generally around \SI{100}{\ampere\per\square\centi\meter}~\cite{krizan_electrical_2026, wang_fabrication_2015}. 

\section{JJ Chemistry and Interface Characterization}

\begin{figure}
    \centering
    \includegraphics[width=\linewidth]{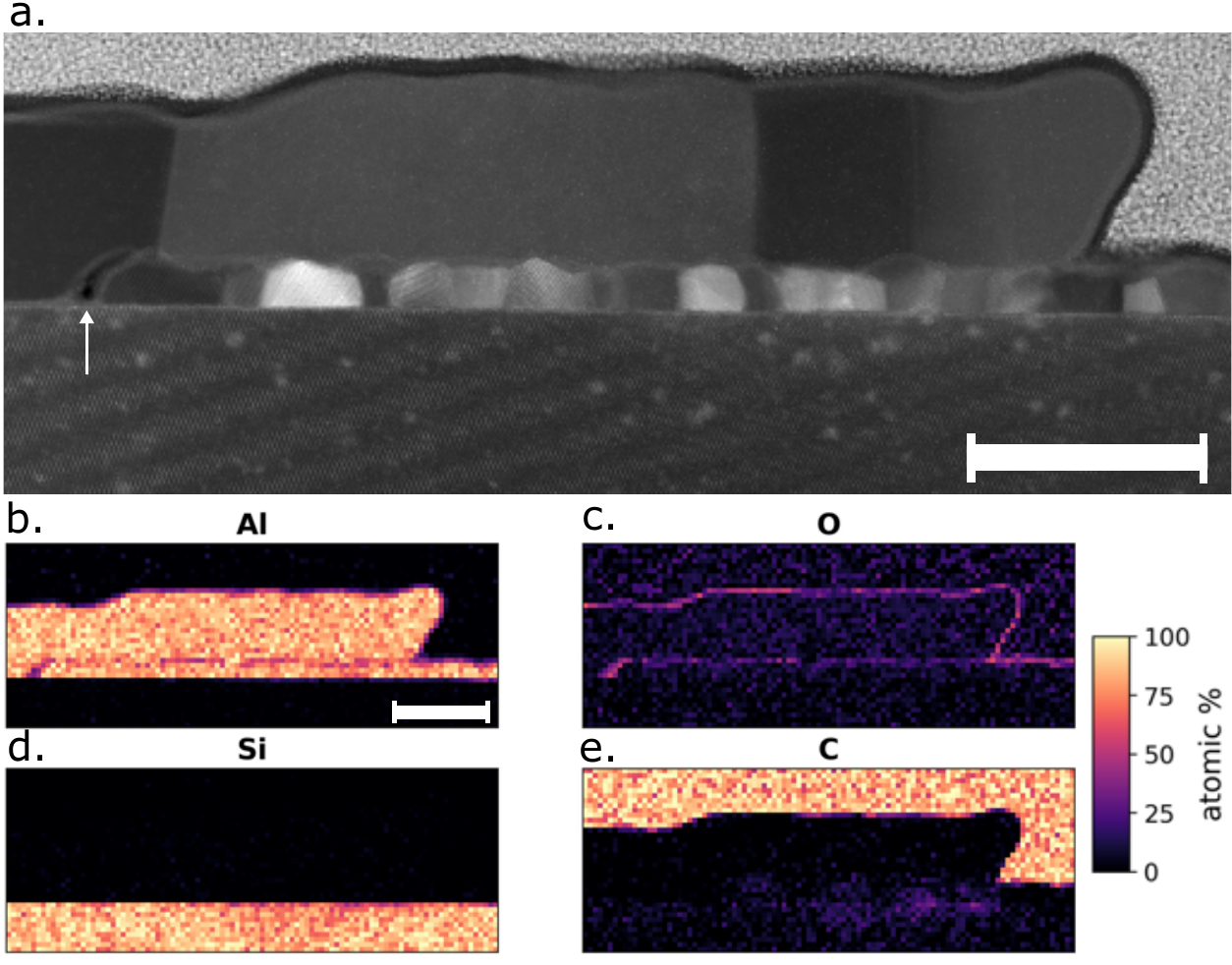}
    \caption{STEM/EELS. (a) Virtual annular dark field image constructed from 4D STEM data showing Al electrode microstructure in the overlap region. 
    The void formed between the side of the bottom electrode and the climbing top electrode is indicated with a white arrow.
    (b)-(e) Quantitative EELS elemental maps of the same region. Diffused carbon signal near the interface is due to electron beam-induced hydrocarbon build-up during previous high-resolution imaging. 
    No carbon is seen specifically concentrated at the Si/Al interface as would be expected for resist residue.
    All scale bars are \SI{100}{\nano\meter}.
    }
     \label{fig:TEM}
\end{figure}

We turn to examination of the chemistry and structure of the JJ at the atomic scale. 
We conducted cross-sectional scanning transmission electron microscopy (STEM) and electron energy loss spectroscopy (EELS) on the device. 
The virtual annular dark field image from the 4D STEM data of the JJ, depicted in Fig.~\ref{fig:TEM}(a), shows clear crystalline Al grains in both the top and bottom electrode.
Grain sizes in the bottom electrode range between 10 and \SI{50}{\nano\meter}, while the grain sizes in the top electrode are between 50 and 100~\si{\nano\meter}, with the largest grain being around \SI{300}{\nano\meter}.
The crystallinity and grain size is similar to ones measured in Josephson junctions elsewhere~\cite{zeng_direct_2015,wolff_structural_2026}.
The oxide layer is clearly visible with a thickness between 1.6 and 1.9~\si{\nano\meter} (see Appendix~\ref{app:TEM}).

The EELS composition map, shown in Fig.~\ref{fig:TEM}(b)-(e) examines the chemical profile within the JJ.
We see that there is no evident carbon contamination within the device, either within the evaporated aluminum or any of the interfaces.
The small carbon signal observed in the Si substrate is attributed to the electron beam induced hydrocarbon build-up that occurred during high-resolution imaging for prior images.
There is also no oxidized silicon found on the substrate under the junction.
This highlights the high cleanliness of the substrate prior to deposition, and the minimization of any residue at the Si/Al interface.
Thus, the trench-based JJs contrasts with resist-based junctions, which may exhibit non-negligible C and O signals at the interface~\cite{gingras_improving_2025}.

We note that there is a void between the side of the bottom electrode and the climbing top electrode (indicated by a white arrow in Fig.~\ref{fig:TEM}(a)).
This is due to the deposition angle of the top electrode counter to the edge of the bottom electrode (see Fig.~\ref{fig:JJfab_SEM}(b)) which results in a shadow, similar to that seen on the non-climbing edge of Manhattan-style junctions. 
We believe that the mobility of the Al atoms during deposition while the surface remains unoxidized avoids a full shadowing: the films are attracted to each other and the lower film appears to `bend' backward during the growth to make contact to the upper film. 

\section{Qubit fabrication and measurement}

\begin{figure*}
    \centering
    \includegraphics[width=\linewidth]{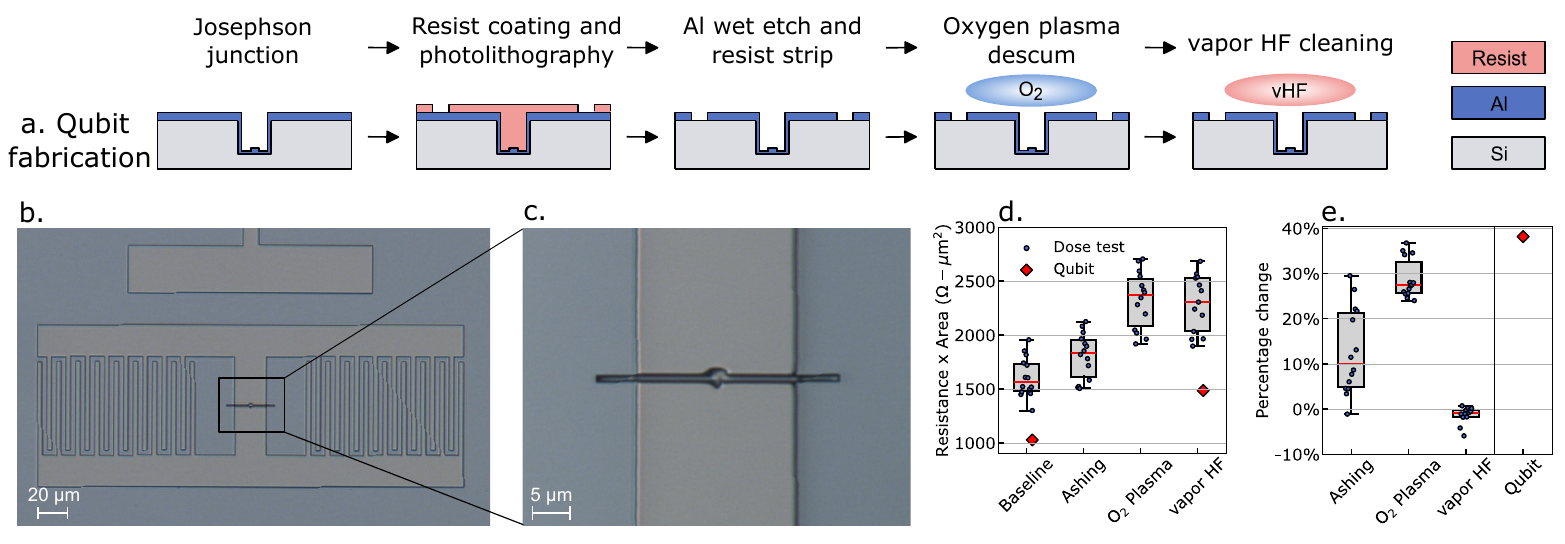}
    \caption{Qubit fabrication. 
    (a) Full qubit fabrication process. 
    The evaporated junction film is coated in photoresist, followed by photolithography and an Al wet etch. 
    The resist is stripped from the chip, with a subsequent oxygen plasma descum and a gentle vapor HF cleaning.
    (b) Optical microscope image of a interdigitated capacitor qubit.
    (c) Magnified image of the JJ location on the qubit.
    (d) Room temperature junction resistance variation on a test chip after four consecutive fabrication steps: after the wet etch process and resist strip (`baseline'), after ashing, after the oxygen plasma step, and after vapor HF treatment (`vapor HF').
    Red diamond indicates a test JJ on the qubit chip that was measured after wet etch and resist strip, and after completion of the full fabrication as detailed in (a).
    (e) Relative percentage increase in resistance from each process step.
    We also indicate the percentage change in resistance of the qubit throughout the entire process (baseline $\rightarrow$ qubit).
    }
    \label{fig:QubitFab}
\end{figure*}

Qubits are patterned onto the chips using the Al-AlO\textsubscript{x}-Al films deposited for the junctions as the film for all features, i.e., the JJ deposition step is the only deposition in the entire fabrication process.
The fabrication process is detailed in Fig.~\ref{fig:QubitFab}(a) (see Appendix~\ref{app:qubitfab}).
The trilayer films are coated in photoresist, followed by patterning in a maskless aligner and a wet etch in aluminum etchant type A. 
The resist is then stripped from the film, followed by an oxygen plasma descum and a gentle vapor HF cleaning process.
The vapor HF process assists with removing hydrocarbons on the surface while also removing the SiO\textsubscript{2} exposed at the bottom of the trench.
Fig.~\ref{fig:QubitFab}(b) and (c) shows an optical microscope image of an interdigitated capacitor qubit post-fabrication with a capacitor gap of \SI{5}{\micro\meter}.

We monitored the resistance of the junction after the oxygen plasma and vapor HF processing, which is depicted in Fig.~\ref{fig:QubitFab}(d).
Sequential processing steps are conducted on a test chip which contains junctions of various areas post-Al wet etch and strip (indicated as `Baseline').
We first test a dedicated oxygen plasma asher (labeled `Ashing'), which results in a small increase in resistances.
However, this increase is non-uniform, with a large variation over the different junction sizes (see Fig.~\ref{fig:QubitFab}(e) for percentage increase per junction by each process).
The subsequent oxygen plasma descum in an Oxford reactive ion etching (RIE) system further increases the resistance (labeled `O\textsubscript{2} Plasma').
In this instance, the resistance increase is significantly more consistent, around 25\%-35\%, in contrast to the asher, which is better for reproducibility.
Finally, the vapor HF process (labeled `vapor HF') does not appreciably affect the resistance, with the percentage being near zero.
This indicates that the process is gentle enough to not damage the junction while etching the SiO\textsubscript{2} and removing any surface hydrocarbons.
We also conduct resonator measurements of Al samples with and without vapor HF, which is shown in Appendix Fig.~\ref{fig:app_res}.

\begin{figure}
    \centering
    \includegraphics[width=\linewidth]{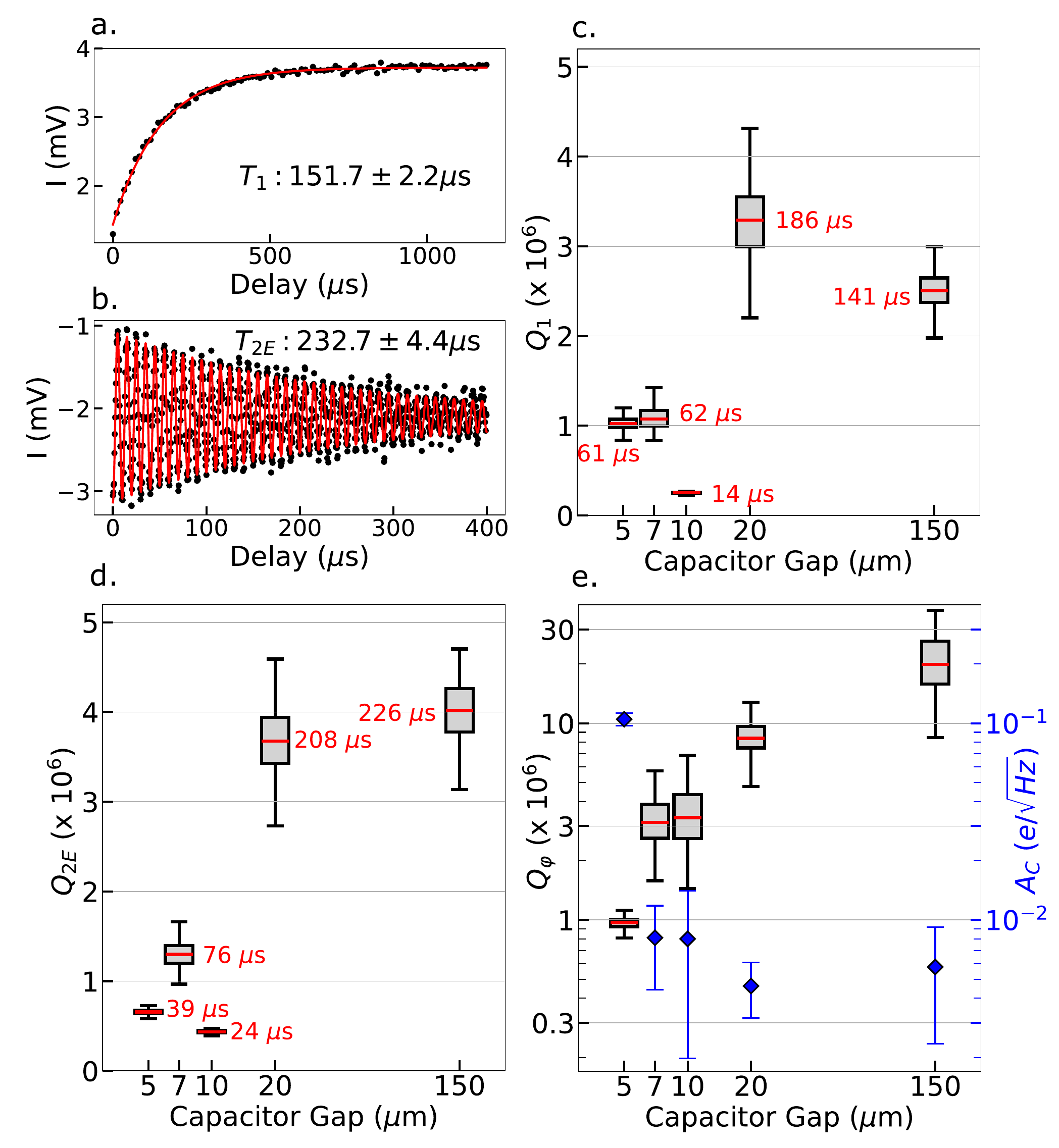}
    \caption{Qubit measurement data. 
    (a), (b) Representative qubit relaxation time $T_1$ and Hahn-echo dephasing time $T_{2E}$ from a qubit with a capacitor gap of \SI{150}{\micro\meter}.
    (c),(d) Qubit energy relaxation quality factor $Q_1$ and Hahn-echo dephasing quality factor $Q_{2E}$ box plots as a function of capacitor gap.
    Median $T_1$ and $T_{2E}$ are indicated in red, with the interquartile range indicated with the box plot and 90th quartile with whiskers.
    (e) Qubit pure dephasing quality factor $Q_\varphi$ box plot (left axis) and calculated upper bound of charge noise amplitude $A_C$ (right axis, blue diamonds) as a function of capacitor gap. 
    }
    \label{fig:QubitData}
\end{figure}

We characterize five qubits on the chip, with the full characterization results shown in Table~\ref{tab:QubitData} (Appendix~\ref{app:Qubit}). 
The qubits have frequencies near \SI{2.8}{\giga\hertz}, and show relaxation times ($T_1$) and Hahn-echo dephasing times ($T_{2E}$) that depend on the gap between the capacitor pads.
Fig.~\ref{fig:QubitData}(a) and (b) show representative measurements of the $T_1$ and $T_{2E}$ data from the qubit with a \SI{150}{\micro\meter} capacitor gap. 
Fig.~\ref{fig:QubitData}(c) and (d) show the qubit energy relaxation and Hahn-echo dephasing quality factors, $Q_{1}=\omega_{\rm qubit}T_1$ and $Q_{2E}=\omega_{\rm qubit}T_{2E}$, respectively, collected in an interleaved manner over several hours in \SI{10}{\minute} intervals. 
Fig.~\ref{fig:QubitData}(e) shows our estimate for the pure dephasing quality factor, $Q_{\varphi} = (1/Q_{2E} - 1/2Q_{1})^{-1}$ from consecutive $T_1$ and $T_{2E}$ measurements in the interleaved dataset.
We present all three coherence metrics as quality factors for ease of direct comparison ($\omega_{\rm qubit}$ across the 5 qubits varies by $\approx6.6\%$, see Appendix Table~\ref{tab:QubitData}).

Figures~\ref{fig:QubitData}(c)-(d) show that the qubit quality factors are sensitive to the capacitor gap size up to a maximum median $Q_1$ of \SI{3.26}{million}. 
This indicates that surface losses dominate the time-averaged performance for smaller gap sizes, scaling with the decrease in surface participation of the electric fields up to some threshold~\cite{wang_surface_2015,gambetta_investigating_2017}.
The \SI{150}{\micro\meter} qubit has a lower $T_1$ and $Q_1$ relative to the \SI{20}{\micro\meter} qubit, which might be related to coupling to box modes or walls due to the larger extent of the electric fields~\cite{houck_controlling_2008}.
One of the qubits measured, namely the \SI{10}{\micro\meter} gap qubit, does not follow the trend, and instead has low energy relaxation times for undetermined reasons.

The trend of the dephasing is also notable (Fig.~\ref{fig:QubitData}(e), left axis). 
We find that $Q_\varphi$ continues to increase as the capacitor gap size increases. 
The low dispersive shifts and readout resonator linewidths suggest that dephasing is not yet limited by shot-noise dephasing, and the transmons all have comparable $E_J/E_C \sim 40-45$ (see Supplementary Table~\ref{tab:QubitData}).
The ratio is lower than typical values ($E_J/E_C>50$), and therefore the $Q_\varphi$ trend could be due to the effects of charge noise due to the higher sensitivity, alongside an effect due to the variation in $E_J/E_C$.
We infer an upper bound to the charge noise amplitude $A_C$ due to charge dispersion (Fig.~\ref{fig:QubitData}(e), right axis) by assuming that the entire $T_\varphi$ contribution is charge dispersion limited (Appendix~\ref{app:dephasing}).
The calculated $A_C$ for four of the qubits is generally within margins of error, and are higher than the $10^{-4}-10^{-3}$~\si{e\per\sqrt{\hertz}} quoted for transmons~\cite{krantz_quantum_2019}.
The smallest capacitor gap qubit has a significantly higher $A_C$, well above typical charge noise values.
The calculation indicates the need for further analysis of pure dephasing to decouple other effects, such as quasiparticles~\cite{kurilovich_correlated_2026,antonenko_effect_2025} and spectrally diffusing two-level systems (TLSs)~\cite{klimov_fluctuations_2018,matityahu_qubit_2024}, from charge dispersion.
We leave this subject subject to future study.

Understanding and quantifying $T_1$ fluctuations is critical to moving towards scalable quantum computers~\cite{berritta_realtime_2026,burnett_decoherence_2019}. 
Therefore, conducting statistical analyses of long time traces is essential to characterizing qubit performance.
These measurements were done in a second cooldown in an alternate dilution refrigerator (see Appendix Fig.~\ref{fig:app_wiring} for the fridge wiring diagrams), with the sample remaining packaged and stored in a laboratory environment for 2 weeks.
The second cooldown included a Traveling Wave Parametric Amplifier (TWPA) that enables a higher signal-to-noise ratio and consequently shorter averaging times.

\begin{figure}[b]
    \centering
    \includegraphics[width=\linewidth]{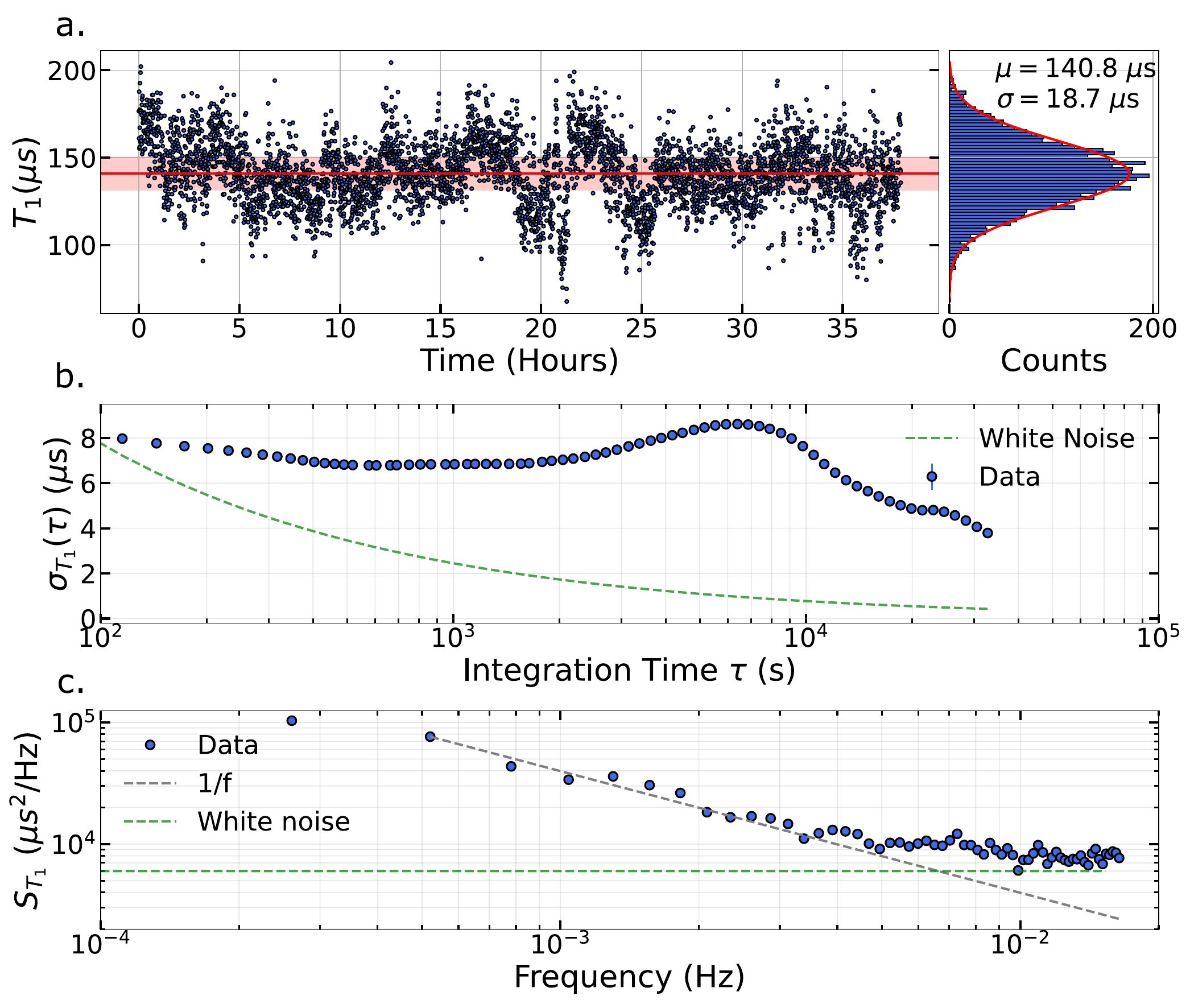}
    \caption{$T_1$ fluctuation analysis of the \SI{150}{\micro\meter} gap qubit from cooldown 2. 
    (a) Fluctuation data taken over \SI{36}{\hour} in \SI{30}{\second} intervals. 
    Histogram showing normal distribution with a mean of $140.8\pm18.7$~\si{\micro\second}. 
    (b) Overlapping Allan deviation plot.
    (c) Spectral density calculated using the Welch method.
    Note that dashed lines indicate the shape of the respective noise within each analysis and are purely illustrative, not fits.
    }
    \label{fig:fluctuations}
\end{figure}

Fig.~\ref{fig:fluctuations}(a) shows the fluctuations of the \SI{150}{\micro\meter} gap qubit over \SI{36}{\hour}, measured in \SI{30}{\second} intervals. 
The $T_1$ fluctuation data of the other qubits are shown in Appendix Fig.~\ref{fig:app_T1fluctuations}.
The $T_1$ fluctuations show distinct drops in performance at various points throughout the time trace, particularly at the \SI{20}{\hour}, \SI{25}{\hour}, and \SI{36}{\hour} mark.
Regardless, the drops are sufficiently infrequent and shallow, resulting in the histogram following a well-defined normal distribution, with a mean $T_1$ of $140.8\pm18.7$~\si{\micro\second}.
This is relatively close to the first cooldown, which signifies reproducible performance over different cooldowns and different dilution refrigerators.

Similar to Burnett et al.~\cite{burnett_decoherence_2019} and Berritta et al.~\cite{berritta_realtime_2026}, we conduct statistical analyses, specifically the overlapping Allan deviation~\cite{riley_handbook_2008} and the spectral density, of the $T_1$ fluctuations.
The overlapping Allan deviation (Fig.~\ref{fig:fluctuations}(b)) shows a relatively clear plateau until the peak around $\tau =10^4$~\si{\second}. 
We note that the Allan deviation is around \SI{7.5}{\micro\second}, significantly lower than the value measured in a qubit which uses a resist-based JJ with a comparable $T_1$ in~\cite{berritta_realtime_2026} (mean $T_1$$\approx$ \SI{168}{\micro\second}, with lowest $\sigma_{T1}\approx13$~\si{\micro\second}), signifying notably lower fluctuations within the device.
The peak at $\tau=10^4$~\si{\second} is likely due to a Lorentzian noise process, since no power-law noise processes produce a peak~\cite{berritta_realtime_2026,riley_handbook_2008}.
Due to the long-range nature of the correlation and relatively shallow increase, this peak could be due to some weakly-coupled TLSs. 
The origin of this peak is difficult to conclude at this time.

The power spectral density (PSD) fluctuation is extracted using the Welch method~\cite{welch_use_1967}, and shown in Fig.~\ref{fig:fluctuations}(c). 
The trend corroborates the data from the Allan deviation, showing that the noise is primarily $1/f$-like, with the saturation at higher frequencies attributed to reaching a white noise floor. 
The lack of any significant deviation from the $1/f$ slope within the spectral density indicates that there is no single, strongly-coupled TLS that is switching to near the qubit frequency within the range of noise frequencies scanned.
This lack of a single significant switching event is true for all five qubits.
Remarkably, we find that $T_{2E}$ fluctuations follow closer to a white noise distribution. 
More details on the fluctuation analysis are contained in Appendix~\ref{app:fluctuations}.

\section{Discussion}

The quality factors of the qubits fabricated using our scheme, while not yet reaching the current state-of-the-art (SOTA)~\cite{bland_millisecond_2025,olszewski_kryptonsputtered_2026}, are nevertheless comparable to earlier iterations of resist-based qubit fabrication methods~\cite{ezratty_there_2023,klimov_fluctuations_2018}.
With the integration of previously established fabrication improvements, the design has the potential to approach SOTA performance.
Moreover, we observe higher quality factors than seen with other novel junction fabrication schemes demonstrated so far~\cite{anferov_improved_2024,tsioutsios_freestanding_2020,hanna_onchip_2026,verjauw_path_2022}.

The $T_1$ fluctuations of the qubits measured here are lower than one would expect for a comparable qubit $T_1$~\cite{berritta_realtime_2026,abdisatarov_demonstrating_2025}, as calculated by the Allan deviation.
To compare against a larger number of works, we utilize the robust coefficient of variation, RCV (defined as the ratio of the interquartile range to the median), as a useful yet imperfect metric for comparison~\cite{botta-dukat_quartile_2023}.
Our RCV remain below 0.2, while other works are usually 0.2-0.3 and beyond~\cite{gingras_improving_2025,biznarova_mitigation_2024,olszewski_kryptonsputtered_2026,colaozanuz_mitigating_2025} (see Appendix Fig.~\ref{fig:app_comparison}).
Such long-term stability of qubit coherence is desirable for larger-scale qubit chips, improving reproducibility and reducing the need for frequent qubit re-characterization. 

There are several avenues available to further improve the fabrication of the qubits shown here.
This involves exploring alternative etch processes, both for the silicon trench and the aluminum film, and further cleaning processes.
Additionally, the lack of an over-etch of the Si due to the use of an Al wet etch makes these devices unable to leverage the reduction in the participation of the lossy interfaces, which could further improve qubit performance, particularly for the more compact designs~\cite{gambetta_investigating_2017,murray_analytical_2020,murray_material_2021}. 
Therefore, the performance shown here is primarily for initial benchmarking of the process, indicating proof-of-principle demonstration of the viability of the scheme in producing high-quality qubits.

A significant consideration in the design and integration of the new JJ design is the ability to integrate the trench-based junction into any existing Si-based planar qubit fabrication process.
The specifics of the design allow for the Si to be etched post-base layer deposition, followed by deposition of the JJ and subsequent etch of the non-JJ film.
The simplicity of the incorporation enables the use of this design in conjunction with improvements in base layer fabrication improvements that have been found in recent years. 

A key feature of the junction design described here is the substantial widening of the process window for JJ fabrication.
The design is significantly more material-agnostic in comparison to resist-based junction fabrication schemes, only limited by the ability of a given material to be directionally evaporated.
The material-agnosticism opens the door to an extensive exploration of the interplay between alternative materials with qubit performance, without the limitation of the damage seen in traditional trilayer junction fabrication schemes.

Overall, this fabrication scheme provides new avenues of performance and materials investigation that was previously unexplored due to limitations of existing processes.
The design shows promise for integration to existing qubit fabrication processes, while removing the hindrances of resist-based and other fabrication methods.
We anticipate future studies involving approaches to substrate annealing and cleaning, integration with high-performing base layers~\cite{bland_millisecond_2025,olszewski_kryptonsputtered_2026}, and detailed noise spectroscopy~\cite{christensen_anomalous_2019}, all in conjunction with alternative JJ materials to explore new avenues to improve qubit performance.

\section{Conclusion}

In conclusion, we have developed a novel Josephson junction fabrication scheme that is more material-agnostic and offers substantially wider thermal and chemical process windows due to avoiding the use of a polymer liftoff mask.
The method is scalable due to the use of standard silicon etching, and integrable with existing semiconductor and superconductor process technology alongside Josephson junction shadow-evaporation tools for wafer sizes up to \SI{200}{\milli\meter}.
We fabricate qubits using this scheme and measure qubits with the highest median $T_1$ of \SI{184}{\micro\second} and $Q_1$ of 3.26 million.
We see minimal interface contamination and find capacitor gap-dependent performance, indicating a surface-participation-ratio limited device which may be improved further through additional post-processing.
Coherence fluctuation analysis using Allan deviation and PSD show that the deviation in the performance is lower compared to other comparable qubits.
Future work will involve the use of alternative electrode and tunnel barrier materials, integration with non-aluminum base layers, and exploration of cleaner fabrication processes and thermal annealing steps.

\section*{Data Availability}

The data that supports this article is openly available at https://doi.org/10.5281/zenodo.19303492~\cite{banerjee_dataset_2026}.

\section*{Author contributions}

TB designed, fabricated, measured, and analyzed the data from all the junction and qubit devices, under the supervision of VF.
SR assisted with the qubit measurements. 
SDF created the FIB samples and conducted all STEM/EELS measurements and analysis, with JJC's supervision.
TB and VF wrote the manuscript with input from all co-authors. 
VF conceived the JJ fabrication scheme and supervised the project. 

\begin{acknowledgments}
We thank Zachary L. Parrott, Anthony P. McFadden and Corey Rae H. McRae (NIST-Boulder) for sharing their transmon qubit design, which we adapted for this study.
We also thank Zachary L. Parrott for his comments on the manuscript and Christopher J. K. Richardson (LPS) for discussions. 
We thank Cornell NanoScale Facility staff for their advice regarding fabrication, namely Tom Pennell, Jeremy Clark, Aaron Windsor, Xinyi Du and Michael Skvarla.
We thank Simon Reinhardt for assistance setting up decoherence analysis.

TB acknowledges funding from the LQC National Quantum Fellowship Program.
SR acknowledges funding from the NSF under award number PHY 2512537. 
This research was supported in part by an appointment to the Department of Defense (DOD) Research Participation Program administered by the Oak Ridge Institute for Science and Education (ORISE) through an interagency agreement between the U.S. Department of Energy (DOE) and the DOD. ORISE is managed by ORAU under DOE contract number DE-SC0014664. All opinions are the author's and do not necessarily reflect the policies and views of DOD, DOE, or ORAU/ORISE.

This work was performed in part at the Cornell NanoScale Facility, a member of the National Nanotechnology Coordinated Infrastructure (NNCI), which is supported by the National Science Foundation (Grant NNCI-2025233).
This work made use of the Cornell Center for Materials Research shared instrumentation facility, Helios FIB, and Thermo Fisher Spectra 300 STEMs whose acquisitions were partly supported by the National Science Foundation (NSF) DMR-2039380 and the Platform for the Accelerated Realization, Analysis, and Discovery of Interface Materials (PARADIM) Award No. DMR-2039380. STEM characterization was supported by DOE DE-SC0023905.
\end{acknowledgments}

\section*{Competing Interests}

The authors declare no competing interests. 

\clearpage
\appendix

\renewcommand{\thefigure}{S\arabic{figure}}
\renewcommand{\thetable}{S\arabic{table}}

\setcounter{section}{0}
\setcounter{figure}{0}
\setcounter{table}{0}
\onecolumngrid
\begin{center}
    \textbf{\Large Supplementary Material: Resist-free shadow deposition using silicon trenches for Josephson junctions in
superconducting qubits} \\
    \vspace{0.5cm}
    Tathagata Banerjee$^{1}$, Stephen Daniel Funni$^{2}$, Saswata Roy$^{3}$, Judy J. Cha$^{2}$, and Valla Fatemi$^{1}$ \\
    \vspace{0.2cm}
    \small $^{1}$School of Applied and Engineering Physics, Cornell University, Ithaca, NY 14853, USA \\
    $^{2}$Department of Materials Science and Engineering, Cornell University, Ithaca, NY 14853, USA \\
    $^{3}$Department of Physics, Cornell University, Ithaca, NY 14853, USA
\end{center}

\twocolumngrid

\section{Device fabrication}\label{app:fabrication}
\subsection{Trench fabrication}\label{app:trenchfab}

We begin by taking an intrinsic, high resistivity ($>10,000$ $\Omega$-cm) undoped, single-side-polished, float zone four-inch Si (100) wafer sourced from WaferPro, and oxidize using a dry oxidation process (no HCl) at \SI{1000}{\celsius} for \SI{20}{\minute} to grow approximately \SI{170}{\nano\meter} of SiO\textsubscript{2} on the wafer.
The oxidized wafer is then spin-coated at \SI{3000}{rpm} with Kayaku 495k PMMA A8 e-beam resist and baked at \SI{170}{\celsius} for \SI{15}{\minute}. 
The measured thickness of the resist is $378\pm4$~\si{\nano\meter}.
The wafer is then spin-coated with DisCharge anti-charging agent from DisChem Inc. at \SI{2000}{rpm} to prevent pattern distortion due to charge accumulation in the insulating SiO\textsubscript{2} during lithography.

Electron-beam lithography is conducted in a JEOL 6300 series system with a beam current of \SI{10}{\nano\ampere} with a dose of \SI{1000}{\micro\coulomb\per\square\centi\metre}.
Post lithography, the wafer is rinsed in deionized water (DI H\textsubscript{2}O) to remove the DisCharge coating.
The pattern is developed at room temperature in a 1:3 MIBK:IPA (methyl isobutylketone:isopropanol) for \SI{1}{\minute}, followed by a \SI{30}{\second} rinse in IPA and blow drying with nitrogen.

SiO\textsubscript{2} etching is conducted in an Oxford PlasmaLab 100 Inductively Coupled Plasma (ICP) System using a CHF\textsubscript{3}/O\textsubscript{2} based process for \SI{1.5}{\minute} (CHF\textsubscript{3}: \SI{52}{sccm}, O\textsubscript{2}: \SI{2}{sccm}, ICP: \SI{2500}{\watt}, RIE: \SI{15}{\watt}, pressure: \SI{5}{mTorr}).
The chamber was prepared with a \SI{10}{\minute} oxygen plasma clean and a \SI{5}{\minute} chamber seasoning with a blank Si wafer prior to etching. 
The e-beam resist is then stripped from the surface in an Integrated Micro Materials AZ300T stripper bath at \SI{70}{\celsius} for \SI{15}{\minute}.
This is followed by two consecutive sonication baths in IPA for \SI{15}{\minute} and \SI{5}{\minute} respectively.

The Si is etched with the SiO\textsubscript{2} hard mask pattern in an HBr-based Oxford PlasmaPro NGP100 ICP-RIE etcher for \SI{3.5}{\minute} (HBr: \SI{20}{sccm}, ICP: \SI{2000}{\watt}, RIE: \SI{40}{\watt}, pressure: \SI{8}{mTorr}).
The etcher was prepared with a \SI{10}{\minute} SF\textsubscript{6} plasma clean and a \SI{10}{\minute} chamber seasoning with a blank Si wafer prior to etching.
Following the etch, the measured thickness of the remaining hard mask is $97\pm4$~\si{\nano\meter}, with a Si trench depth of approximately \SI{1.1}{\micro\metre} and a sidewall angle of approximately \SI{85}{\degree}-\SI{88}{\degree}. 
The remaining SiO\textsubscript{2} hard mask is stripped by soaking in a room temperature 10:1 hydrofluoric acid (HF) bath (Transene Company, Inc.) for \SI{5}{\minute} followed by a rinse in DI H\textsubscript{2}O for \SI{30}{\second} and blow drying with air. 

For dicing, the wafer is coated with Microposit S1813 at \SI{4000}{rpm} and baked at \SI{90}{\celsius} for \SI{60}{\second}. 
The wafer is then diced into \SI{7.5}{\milli\meter}$\times$\SI{7.5}{\milli\meter} pieces using a DISCO dicing saw, followed by a resist stripping process similar to the one conducted post-SiO\textsubscript{2} etch: the pieces are soaked in an AZ300T stripper bath at \SI{70}{\celsius} bath for \SI{15}{\minute} followed by two consecutive \SI{15}{\minute} IPA sonication baths and blow drying with air. 

\subsection{Josephson junction fabrication}\label{app:JJfab}

Immediately prior to JJ deposition, the etched sample is soaked in a 10:1 buffered oxide etch (BOE) solution (Transene Company, Inc.) for \SI{60}{\second} to remove any remaining resist residue and SiO\textsubscript{2}.
The samples are then rinsed in DI H\textsubscript{2}O for \SI{30}{\second} followed by blow drying with air.

The samples are loaded into an Angstrom Quantum Series electron beam evaporation system, which is equipped with separate evaporation and oxidation chambers. 
While in the loadlock, the oxidation chamber is flushed with \SI{1}{Torr} oxygen for \SI{10}{\second} three consecutive times to flush any contamination from the oxygen lines.
The base pressure prior to evaporation is below $2\times10^{-8}$~\si{Torr}.

We evaporate 5N purity aluminum from a copper hearth without a crucible.
No ion milling is conducted prior to deposition.
The sample is first tilted to an angle of \SI{49}{\degree}, after which Al for the first electrode is deposited at \SI{2}{\angstrom\per\second} until a nominal thickness of \SI{30.5}{\nano\meter}.
The Al is oxidized at \SI{1}{Torr} for \SI{5}{\minute} using $100\%$ O\textsubscript{2}, followed by the deposition of the second electrode at a \SI{-49}{\degree} angle.
The second electrode is also evaporated at \SI{2}{\angstrom\per\second} until it reaches a nominal thickness of \SI{123}{\nano\meter}.
The true thicknesses of the two electrodes are \SI{20}{\nano\meter} and \SI{80}{\nano\meter} respectively, due to the lower surface flux from the tilted angle.

\subsection{Qubit fabrication}\label{app:qubitfab}

The chip with the evaporated film and the Josephson junctions is spin-coated with HMDS-P20 primer (Transene Company, Inc.) at \SI{2500}{rpm} followed by a coating of Microposit S1813 photoresist at \SI{4000}{rpm}. 
The resist is then baked at \SI{110}{\celsius} for \SI{1}{\minute}.
The pattern is exposed in a Heidelberg Instruments MLA 150 Advanced Maskless Aligner with a dose of \SI{110}{\milli\joule} and a defocus of -2 (\SI{-18}{\micro\meter}).
The pattern is developed in room temperature AZ726 MIF for \SI{60}{\second} followed by a rinse in DI H\textsubscript{2}O for \SI{30}{\second} and blow drying with air.

The pattern is wet-etched into the Al film with a \SI{1.5}{\minute} soak in room temperature Aluminum etchant - type A (Transene Company, Inc.), composed of 40-80~\si{\percent} phosphoric acid, 3-20~\si{\percent} acetic acid and 1-5~\si{\percent} nitric acid.
The wet etch is followed by soaking the sample in DI H\textsubscript{2}O for \SI{30}{\second} and blow drying with air.

The resist is stripped from the chip through subsequent sonication in acetone and IPA for \SI{5}{\minute} each and blow drying with air.
We then remove resist residue using a \SI{1}{\minute} O\textsubscript{2} plasma descum process (\SI{75}{\watt},O\textsubscript{2}: \SI{50}{sccm}, pressure: \SI{60}{mTorr},) in an Oxford PlasmaLab 80+ RIE system.
Finally, the chip undergoes treatment in vapor hydrofluoric acid using a Primaxx $\upmu$Etch system.
The process begins with a \SI{15}{\minute} \textit{in-situ} drying process at \SI{45}{\celsius}, with vaporized ethanol (\SI{300}{sccm}) and nitrogen gas (\SI{1500}{sccm}) at a pressure of \SI{90}{Torr}.
This is immediately followed by the \SI{1}{\minute} vapor HF process (vHF: \SI{190}{sccm}, ethanol: \SI{210}{sccm}, N\textsubscript{2}: \SI{1425}{sccm}).
The chamber is then purged three times with nitrogen before venting.
The chip is then wire bonded, packaged, and loaded into the cryostat for measurement.

\section{Room temperature JJ measurement}\label{app:JJ}

\begin{figure}
    \centering
    \includegraphics[width=\linewidth]{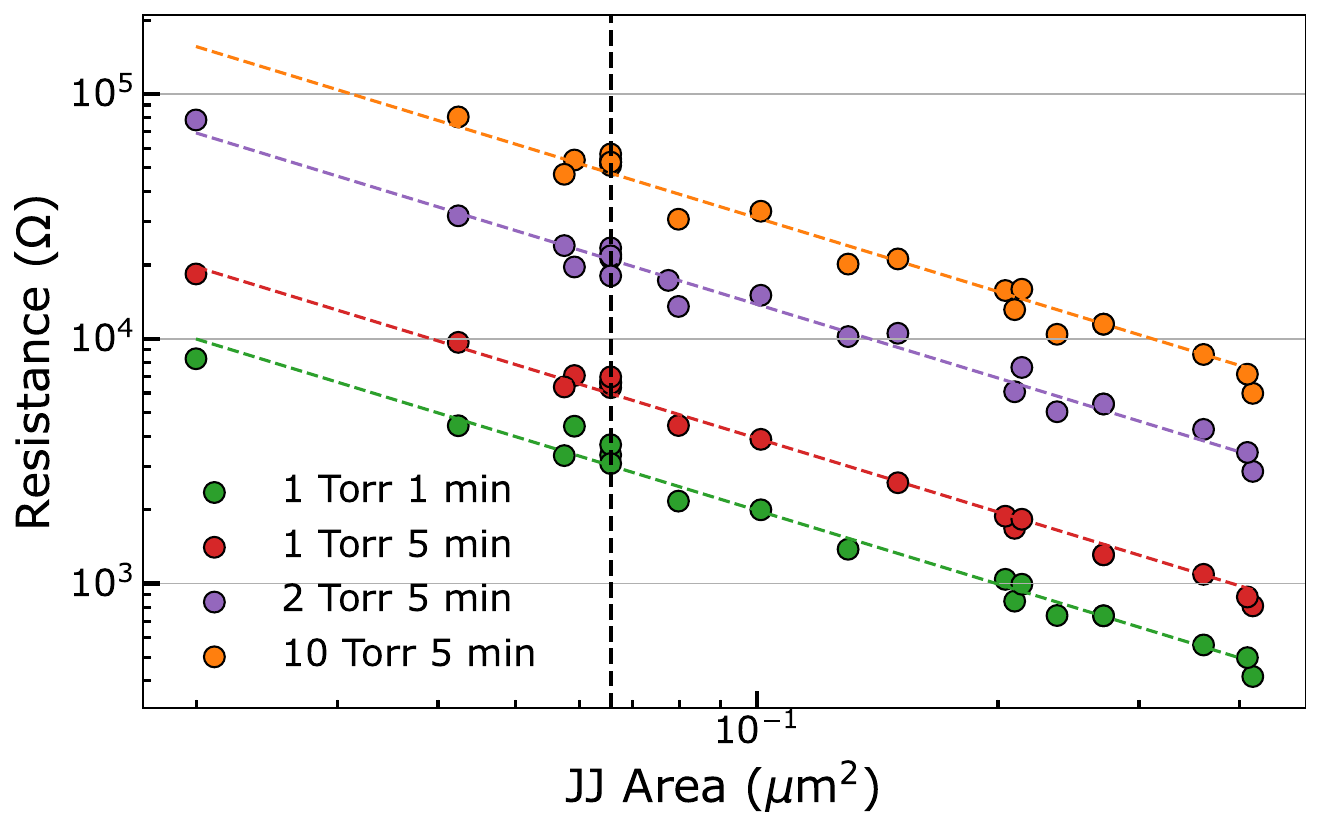}
    \caption{Oxidation dose testing. 
    We conduct four different oxidation doses on test chips containing a variety of JJ areas to examine $J_C$ and the scaling.
    Lines indicate the mean resistance-area product.
    Black line indicates the junction area for the device shown in Fig.~\ref{fig:JJfab_SEM} and the qubits in Figs.~\ref{fig:QubitFab}-\ref{fig:fluctuations}.}
    \label{fig:app_JJ}
\end{figure}

We conduct dose testing of Josephson junctions (JJ) fabricated using the Si-trench scheme with a variety of oxidation doses and areas.
Dose testing data of the JJs are shown in Fig.~\ref{fig:app_JJ}.
Room-temperature two-probe measurements are conducted on all junctions using an Everbeing EB-6 high precision probe station and a Keithley 4200A-SCS Parameter Analyzer.
The measured resistance between probe tips is less than \SI{3}{\ohm}.

We estimate the critical current density $J_c = I_c/A$ with the Ambegaokar-Baratoff formula~\cite{ambegaokar_tunneling_1963},
\begin{equation}
    I_c = \frac{\pi\Delta}{2eR_n},
\end{equation}
where $I_c$ is the critical current, $A$ is the junction area, $R_n$ is the normal state resistance, and we choose $\Delta=$ \SI{180}{\micro\electronvolt} to represent the superconducting energy gap.
Combining the $J_C$ and $I_C$ formula, we find;
\begin{align}
    J_c = \frac{I_cR_n}{R_nA} = \frac{\pi\Delta}{2eR_nA},\\
    J_c = \frac{282.7 \mu\text{eV}}{R_nA},
\end{align}
where $R_nA$ is simply the resistance-area product of the Josephson junction.
The calculated $J_C$ of the various doses is shown in Table~\ref{tab:SI_Dose}. 
The $J_C$ is similar to junctions measured elsewhere~\cite{krizan_electrical_2026}.
The area used for qubit fabrication, \SI{360}{\nano\meter}$\times$\SI{180}{\nano\meter}, is depicted with a dashed black line in Fig.~\ref{fig:app_JJ}.

Fig.~\ref{fig:app_JJ} shows that the resistance-area relation follows the expected trend, $R_n\propto1/A$, across a range of areas, between \SI{0.02}{\square\micro\meter} and \SI{0.42}{\square\micro\meter}.
This indicates that the junction area is stable and the etch pattern does not undergo significant warping or variation as the lateral dimensions increase by a factor of $\approx5$.

\begin{table}
    \centering
    \begin{ruledtabular}
    \begin{tabular}{cccc}
         Pressure & Time & Mean $R_nA$ & $J_C$ \\ \hline
         \SI{1}{Torr} & \SI{1}{\minute} & $199.1\pm24.2$~\si{\ohm\square\micro\meter} & $142.1\pm17.3$~\si{\ampere\per\square\centi\meter}\\
         \SI{1}{Torr} & \SI{5}{\minute} & $391.8\pm33.5$~\si{\ohm\square\micro\meter}& $72.2\pm6.2$~\si{\ampere\per\square\centi\meter}\\
         \SI{2}{Torr} & \SI{5}{\minute} & $1381.6\pm152$~\si{\ohm\square\micro\meter}& $20.5\pm2.3$~\si{\ampere\per\square\centi\meter}\\
         \SI{10}{Torr} & \SI{5}{\minute} & $3114.3\pm382$~\si{\ohm\square\micro\meter}& $9.1\pm1.1$~\si{\ampere\per\square\centi\meter}\\
    \end{tabular}
    \caption{Measured mean resistance-area product $R_nA$ and calculated critical current density $J_C$ of the JJs from oxidation dose testing.}
    \label{tab:SI_Dose}
    \end{ruledtabular}
\end{table}

\section{Further chemical and interface characterization}\label{app:TEM}

Electron transparent cross-sectional samples samples were prepared for STEM with a Thermo-Fisher Helios G4-UX Dual-Beam SEM/FIB with final thinning at \SI{2}{\kilo\volt}. 
A Thermo-Fisher Scientific (TFS) Spectra S/TEM operating at \SI{300}{\kilo\volt} with a \SI{21.4}{\milli\radian} convergence angle and a beam current of \SI{46}{\pico\ampere} was used for all experiments. 
High angle annular dark field (HAADF) imaging was acquired with a detector collection range of 48–200~\si{\milli\radian}. 4D STEM data was collected on an EMPAD direct electron detector using a \SI{1}{\milli\second} dwell time per pixel. 
The virtual annular dark field image was formed from the 4D STEM data set by integrating the intensity in each diffraction pattern over the angular range of 23-35~\si{\milli\radian}, which maximized the diffraction contrast. 

\begin{figure}
    \centering
    \includegraphics[width=\linewidth]{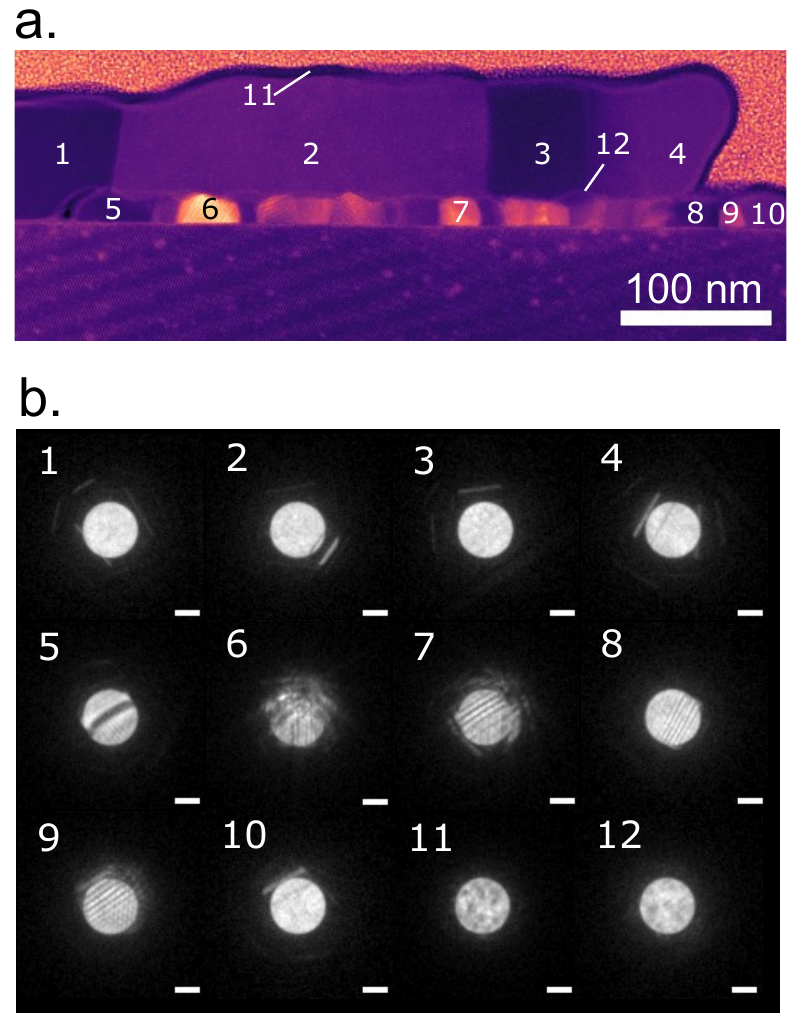}
    \caption{Extended STEM/EELS data. 
    (a) Virtual annular dark field image (same as Fig.~\ref{fig:TEM}(a) but with colormap applied)
    (b) Selected convergent beam electron diffraction (CBED) patterns showing crystallinity of Al electrode grains. 
    Note Kikuchi lines and/or shadow lattice images in 1-10, indicating crystalline order. 
    Patterns 11 and 12 are from amorphous materials: evaporated carbon (deposited prior to FIB sample preparation) and AlO\textsubscript{x} junction barrier, respectively. 
    These two patterns lack crystalline signatures seen in the Al metal electrodes. 
    The mottled contrast in the bright field disk is also indicative of amorphous order. 
    CBED scale bars are \SI{20}{\milli\radian}. }
    \label{fig:app_TEM1}
\end{figure}

EELS data was collected using a TFS Iliad spectrometer with Zebra detector~\cite{lazar_enabling_2025}. 
The EELS acceptance angle was \SI{51.6}{\milli\radian} and the dispersion was \SI{423}{\milli\electronvolt\per channel}. 
All STEM data analyses, including quantitative EELS mapping, were performed using the python package SingleOrigin~\cite{funni_sdfunni_2025}.

Fig.~\ref{fig:app_TEM1}(a) shows the virtual annular dark field image, identical to the one shown in Fig.~\ref{fig:TEM}(a) with a color map applied.
Included are select diffraction patterns from 4D STEM (Fig.~\ref{fig:app_TEM1}(b)).
Crystalline grains show Kikuchi lines and/or shadow lattice images.
We see that all grains except the two amorphous layers -- the AlO\textsubscript{x} tunnel barrier and the evaporated carbon from the FIB preparation -- show either indication of crystallinity.

\begin{figure}
    \centering
    \includegraphics[width=\linewidth]{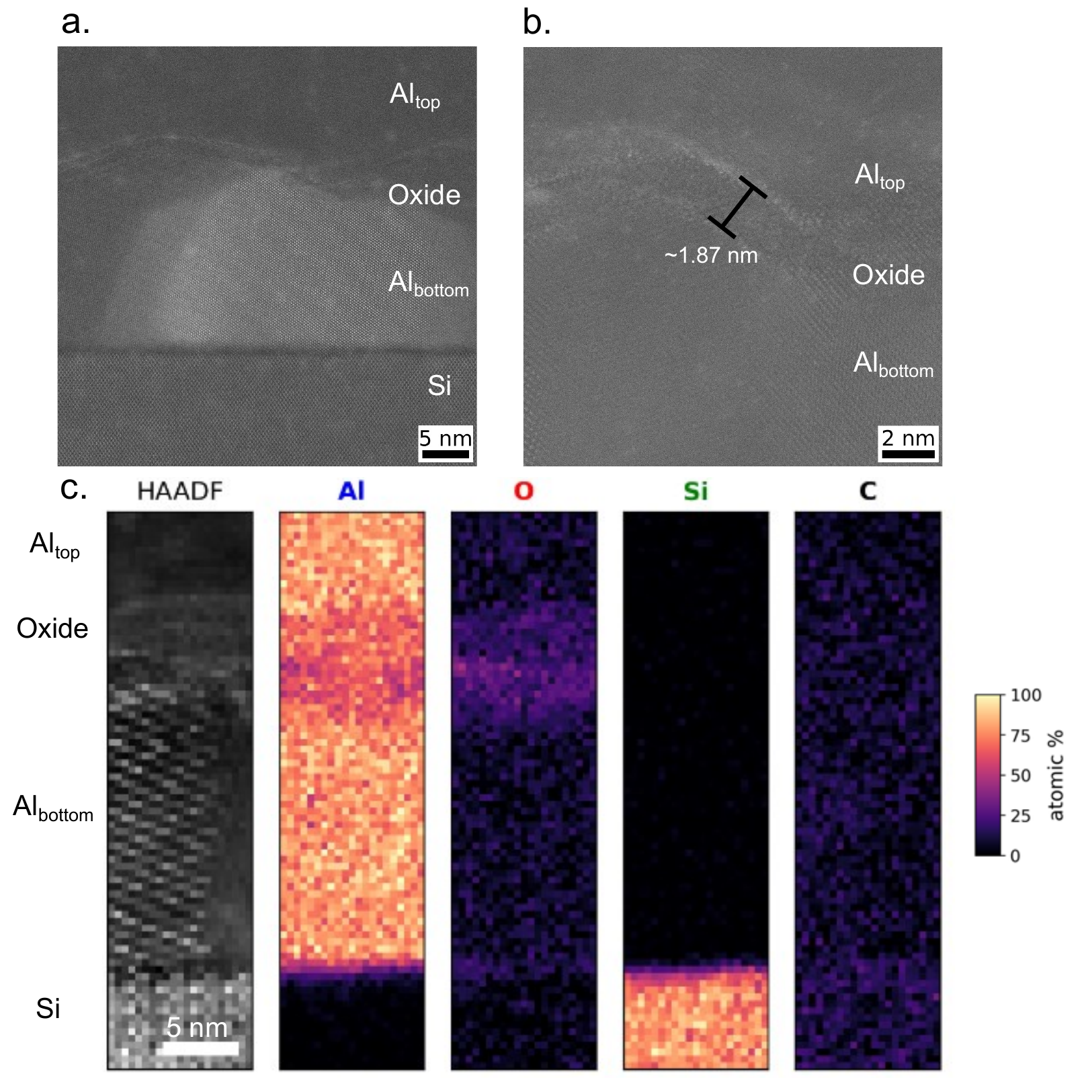}
    \caption{(a) STEM-HAADF image of device interfaces. 
    The near zone axis grain in this image is highly crystalline and corresponds to grain 6 in Fig.~\ref{fig:app_TEM1}. 
    (b) STEM-HAADF image of the oxide layer. 
    (c) Magnified EELS mapping of elemental composition of the oxide barrier and Al/Si interfaces processed the same way as in Fig.~\ref{fig:TEM}(b). 
    Note the lack of carbon contamination at the substrate interface.}
    \label{fig:app_TEM2}
\end{figure}

Fig.~\ref{fig:app_TEM2}(a) and (b) show a zoomed in cross-section. 
The grains and the oxide barrier are clearly visible.
The oxide thickness is estimated to be around ~\SI{1.9}{\nano\meter}. 
Beyond the center of Fig.~\ref{fig:app_TEM2}(b), we see that the oxide barrier seems to diverge.
This is an artifact of the projection and the roughness of the bottom electrode.
It is likely that the apparent divergence is due to the oxide growth on two different grains or the different facets of the same grain.

Fig.~\ref{fig:app_TEM2}(c) shows elemental mapping identical to Fig.~\ref{fig:TEM}(b)-(e), but from a higher resolution EELS data set. 
The enlarged view corroborates the observation that there is minimal carbon contamination at the interfaces or within the film.
The oxygen is localized to the the tunnel barrier, with no oxide present at the Si/Al interface.
The diffused background signal for the carbon and oxygen is a result of the imaging process (hydrocarbon contamination) and sample preparation (air exposure after FIB preparation) respectively.
Any carbon or oxygen intrinsic to the as-prepared device is not detectable in this data below the $\approx5\%$ level due to masking by these artifacts.

\section{Al resonator data}\label{app:resonator}

\begin{figure}
    \centering
    \includegraphics[width=\linewidth]{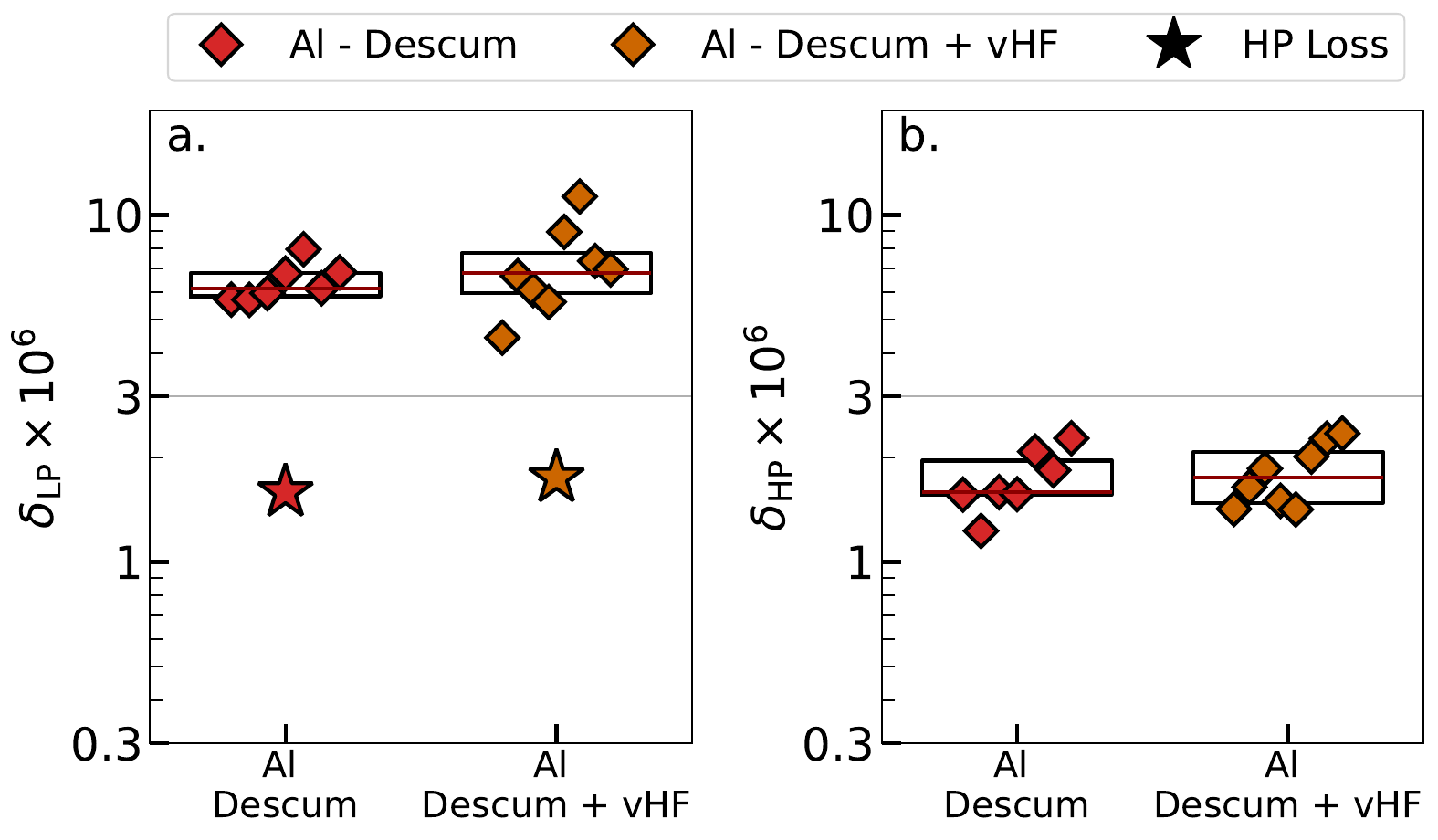}
    \caption{Al resonator data. 
    (a) low power loss ($\delta_{\rm LP})$ and (b) high power loss ($\delta_{\rm HP})$ for an Al resonator processed similarly to the qubit chip, with and without the vapor HF step. 
    We see that the vapor HF does not affect the resonator loss in any meaningful way.}
    \label{fig:app_res}
\end{figure}

We measure Al resonators to examine if the vapor HF impacts performance, shown in Fig.~\ref{fig:app_res}.
The fabrication is similar to the qubit chip, with the primary difference being the use of only a single Al film \SI{60}{\nano\meter} thick, and patterning of the resonators using a g-line stepper. 
The pattern used is from the Boulder Cryogenic Testbed, which contains eight branches on each chip with target frequencies between $4-8$~\si{\giga\hertz}~\cite{kopas_simple_2022}.
We see no measurable difference between the samples with and without the vapor HF process.

\section{Qubit design and measurement}\label{app:Qubit}

\begin{figure}
    \centering
    \includegraphics[width=\linewidth]{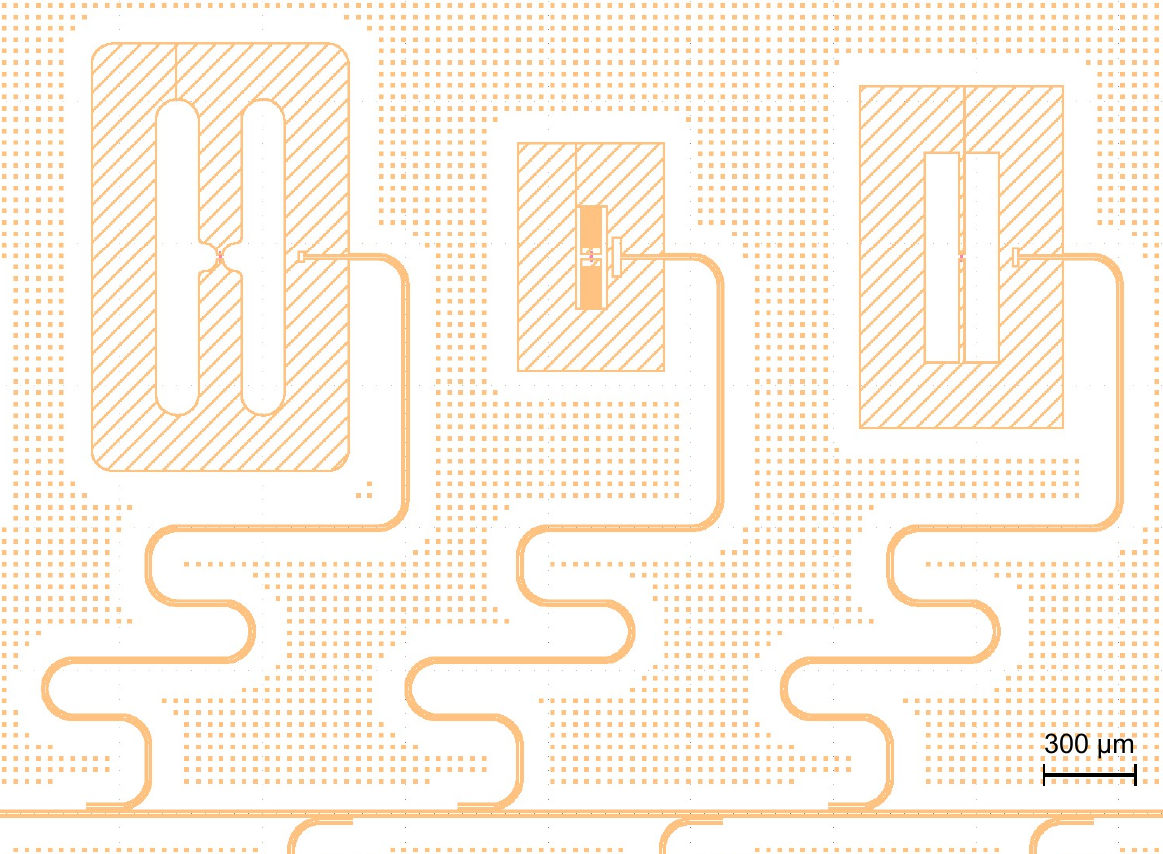}
    \caption{
    GDSII image of as section qubit chip.
    We see the three types of qubit capacitors present in the design -- rectangular pad capacitor (Q-20), interdigitated capacitor (Q-5, Q-7, and Q-10), and paddle capacitor (Q-150). 
    The qubits are driven through the feedline.
    The design contains flux traps to trap stray magnetic vortices.}
    \label{fig:app_chip}
\end{figure}

The qubit design was adapted from a design shared with us by Parrott, McFadden, and McRae at NIST-Boulder.
We show a section of the GDSII of the chip in Fig.~\ref{fig:app_chip}.
The chip contains three different designs of capacitor pads -- rectangular, interdigitated, and paddles.
The paddle capacitor has the largest capacitor gap of \SI{150}{\micro\meter} (Q-150), followed by the rectangular pad capacitor with a gap of \SI{20}{\micro\meter} (Q-20).
The three smaller qubits measured -- 5, 7, and \SI{10}{\micro\meter} gaps (Q-5, Q-10, and Q-20) -- are interdigitated capacitors.

There is no independent drive line for the qubit, instead it is driven through the feedline and the readout resonator. 
There are also patterned square \SI{8}{\micro\meter}$\times$\SI{8}{\micro\meter} flux traps on the chip away from the resonators and capacitors to trap stray vortices. 
The nearest patterned traps are at least \SI{100}{\micro\meter} away from the major patterned features.

The full characterization details of all 5 qubits is shown in Table~\ref{tab:QubitData}. 
The characterization was conducted during the first cooldown.
In the second cooldown, $f_{\rm res}$ changed less than \SI{100}{\kilo\hertz}, qubit frequencies were within \SI{100}{\mega\hertz} of the first cooldown (generally moved lower), and $T_1$ times did not change more than $20\%$. 
We note that the suppression of the coherence time of Q-150 relative to Q-20 has been internally observed in other samples made with standard resist-based JJs and qubits with different base layers, indicating that this is not related to the scheme we have presented here.

\begin{table*}
    \centering
    \begin{ruledtabular}
    \begin{tabular}{cccccc}
            Qubit&Q-5&Q-7& Q-10&Q-20& Q-150\\ \hline 
            Capacitor gap&\SI{5}{\micro\meter}&\SI{7}{\micro\meter}& \SI{10}{\micro\meter} &\SI{20}{\micro\meter} & \SI{150}{\micro\meter} \\ 
          $f_{\rm res}$& \SI{6.62}{\giga\hertz} &\SI{6.73}{\giga\hertz}&\SI{6.80}{\giga\hertz}&\SI{6.65}{\giga\hertz}& \SI{6.78}{\giga\hertz} \\
         $f_{\rm qubit}$& \SI{2661.0}{\mega\hertz}& \SI{2714.2}{\mega\hertz}& \SI{2838.1}{\mega\hertz}&\SI{2811.6}{\mega\hertz}& \SI{2828.0}{\mega\hertz} \\
          $T_1$& $61.24\pm$\SI{4.7}{\micro\second}& $66.42\pm$\SI{14.9}{\micro\second}& $14.07\pm$\SI{0.53}{\micro\second}&$184.79\pm$\SI{27.85}{\micro\second}& $141.18\pm$\SI{11.6}{\micro\second}\\
          $T_{2R}$& $25.10\pm$\SI{5.74}{\micro\second}& $24.98\pm$\SI{23.03}{\micro\second}& $9.33\pm$\SI{3.25}{\micro\second}& $57.28\pm$\SI{26.54}{\micro\second}& $49.06\pm$\SI{23.03}{\micro\second}\\
          $T_{2E}$& $39.08\pm$\SI{1.88}{\micro\second}& $76.06\pm$\SI{9.1}{\micro\second}& $24.35\pm$\SI{1.1} {\micro\second}& $207.15\pm$\SI{24.3} {\micro\second}& $225.75\pm$\SI{18.9} {\micro\second}\\
 $\chi$& \SI{-90}{\kilo\hertz}& \SI{-110}{\kilo\hertz}& \SI{-50}{\kilo\hertz}& \SI{-70}{\kilo\hertz}&\SI{-70}{\kilo\hertz}\\
 $-\alpha$& \SI{168}{\mega\hertz}& \SI{190}{\mega\hertz}& \SI{198}{\mega\hertz}& \SI{190.5}{\mega\hertz}&\SI{177}{\mega\hertz} \\
 $\kappa_{\rm res}$& \SI{181}{kHz}& \SI{206}{kHz}& \SI{204}{kHz}& \SI{667}{kHz}&\SI{278}{kHz}\\ \hline
 $T_\varphi$& $57.66\pm$\SI{4.24}{\micro\second}& $202.21\pm$\SI{83.71}{\micro\second}&$216.80\pm$\SI{139.89}{\micro\second}&$500.85\pm$\SI{149.38}{\micro\second}&$1281.33\pm$\SI{666.32}{\micro\second}\\\
 $E_J/E_C$& 46.85& 39.79& 39.99& 41.87&47.51\\
 $A_C$& $104.83\times10^{-3}$~\si{e\per\sqrt{\hertz}}&$8.12\times10^{-3}$~\si{e\per\sqrt{\hertz}} & $8.01\times10^{-3}$~\si{e\per\sqrt{\hertz}}&$4.62\times10^{-3}$~\si{e\per\sqrt{\hertz}} &$5.77\times10^{-3}$~\si{e\per\sqrt{\hertz}}\\
    \end{tabular}
    \caption{Measured and calculated parameters of the 5 qubits on the chip during the first cooldown.
    Measured parameters: $f_{\rm res}$: resonator frequency, $f_{\rm qubit}$: qubit frequency; $T_1$: qubit relaxation time; $T_{2R}$: Ramsey dephasing time; $T_{2E}$: Hahn-echo dephasing time; $\chi$: dispersive shift; $-\alpha$: anharmonicity;  $\kappa_{\rm res}$: resonator energy decay rate. 
    Calculated parameters: $T_\varphi$: pure dephasing time; $E_J/E_C$: ratio calculated via Hamiltonian numerical diagonalization; $A_C$: charge noise amplitude.
    }
    \label{tab:QubitData}
    \end{ruledtabular}
\end{table*}

\subsection*{Qubit measurement}

The qubit is measured in two different Bluefors LD250 dilution refrigerators, both with base temperatures below \SI{10}{\milli\kelvin}.
The control and the readouts in both cooldowns are conducted using the same Quantum Machines OPX+ instrument, in conjunction with a Quantum Machines Octave instrument.
Control pulses are generated in the OPX+ and up-converted using local oscillators (LO) and IQ mixers present within the Octave.
Pulses are digitally triggered, with a trigger buffer on either side of the pulse.
Measurements were conducted with a readout length of \SI{5}{\micro\second}.
The $\pi$-pulse had a length of \SI{1}{\micro\second} and was shaped like a gaussian, terminated at $5\sigma$.

Relaxation time $T_1$ was measured by sending a $\pi$-pulse followed by a time delay $\tau$ before measurement.
The waiting time between consecutive measurements was set to at least $6T_1$ to ensure the qubit was depopulated before the subsequent measurement.
The resulting decay was by a standard exponential with an offset, $Ae^{\tau/T_1}+C$\cite{krantz_quantum_2019}.

The Hahn-echo measurement to determine $T_{2E}$ was conducted by sending a pulse sequence of $\pi/2 - \pi -\pi/2$ with a $\tau/2$ delay between the pulses with zero detuning from qubit frequency.
The second $\pi/2$-pulse has a phase offset, increasing with each delay time. 
A large phase increment ensures the timescale of the decay and oscillations are well separated.
The resulting oscillations were fit with a convolution of a sinusoid and an exponential decay with a phase offset, $Ae^{-\tau/T_{2E}}\sin(2\pi\Delta\tau+\phi)+C$.

The addition of a TWPA in the second cooldown did not alter any of the sequences. 
The TWPA pump was pulsed on for the duration of the readout pulse and a \SI{100}{\nano\second} buffer on each side of the readout pulse.

\section{Dephasing and charge noise}\label{app:dephasing}

We estimate bounds on the photon-induced dephasing rate $\Gamma^p_\varphi$ of the qubit using the standard formulas~\cite{sears_photon_2012,yan_flux_2016},
\begin{align}
    \Gamma^{p}_\varphi = \bar{n}\kappa_{\rm res} \frac{\chi^2}{\kappa_{\rm res}^2 + \chi^2 }, \\
    Q^{p}_\varphi = \frac{2\pi f_{\rm qubit}}{\Gamma_\varphi},
\end{align}
where $\bar{n}$ is the residual photons in the resonator, $\kappa_{\rm res}$ is the resonator decay rate, $\chi$ is the dispersive shift, and $f_{\rm qubit}$ is the qubit frequency.
If we assume a typical $\bar{n}=0.01$, we find $Q^p_\varphi$ to be lowest for Q-7, about 37.3 million (T$^p_\varphi=$ \SI{2.19}{\milli\second}), and highest for Q-20, at 243 million ($T^p_\varphi=$ \SI{13.76}{\milli\second}). 
This is consistent with the lack of obvious saturation in $Q_\varphi$ in the data. 
Therefore, we infer that the qubits are unlikely to be limited by resonator shot noise dephasing.

We estimate  $E_J/E_C$ by using a root-finding algorithm which uses the energy levels calculated via numerically diagonalizing the Hamiltonian~\cite{koch_chargeinsensitive_2007},
\begin{equation}
    \hat{H} = 4E_C(\hat{n}-n_g)^2-E_J\cos{\hat{\phi}},
\end{equation}
with offset charge $n_g=0.25$ and the experimentally measured $f_{01}$ and $-\alpha$ to determine the corrected $E_J$ and $E_C$.

We estimate charge dispersion $\epsilon_{01}$ using the same numerical diagonalization process with the corrected $E_J$ and $E_C$, to solve the equation~\cite{koch_chargeinsensitive_2007},
\begin{equation}
    \epsilon_{01} = E_{01}(n_g=0.5)-E_{01}(n_g=0),
\end{equation}
which yields values in the $10-50$~\si{\kilo\hertz} range. 
Finally, we use $\epsilon_{01}$ to calculate the charge noise amplitude $A_C$ using the formula~\cite{ithier_decoherence_2005},
\begin{equation}
    A_C = \frac{1}{2\pi^2\sqrt{\ln{2}}}\frac{1}{T_\varphi\epsilon_{01}}.
\end{equation}

$A_C$ is plotted alongside $T_\varphi$ as a function of the capacitor gap in Fig.~\ref{fig:QubitData}(e), with the numbers reported in Table~\ref{tab:QubitData}.
The values of $A_C$ for Q-7, Q-10, Q-20, and Q-150 fall in the range $4-8\times10^{-3}$~\si{e\per\sqrt{\hertz}}, higher than the generally quoted $1\times10^{-3}$~\si{e\per\sqrt{\hertz}}~\cite{krantz_quantum_2019}. 
For the calculation conducted here, we assumed that all of the $T_\varphi$ is attributed to the charge dispersion to examine if the device is charge noise limited primarily due to the $E_J/E_C$.
Therefore, the $A_C$ calculated here can be treated as an upper bound to the $1/f$ charge noise amplitude of the device.
If so, then moving to a design with a higher $E_J/E_C$ ratio will alleviate the limitation, and we should observe higher $T_\varphi$ values in those qubits, irrespective of $T_1$ and $T_{2E}$ values.
We leave a further investigation on the source of dephasing to future work. 

$A_C$ for Q-5, however, is significantly larger than any of the other calculated $A_C$, and higher than $A_C$ measured in significantly lower $E_J/E_C$ qubits~\cite{christensen_anomalous_2019}. 
Q-5, having only \SI{5}{\micro\meter} capacitor gaps in an interdigitated design, is significantly more sensitive to the losses on the surface of the pads (mutual capacitance $C_{12}/C_\Sigma\approx0.85$).
The calculated upper bound to the charge noise amplitude is therefore likely well above the actual value. 

The trend in the $A_C$ does not seem to fully capture the relative increases in $T_\varphi$. 
The increase in $E_J/E_C$ and the capacitor gap would also result in a decreasing sensitivity to TLS-related charge fluctuations within the capacitor pads as well (due to reduced mutual capacitance), which would improve the pure dephasing of the qubits. 
Thus, it is likely that it is both the charge dispersion, TLS sensitivity, and quasiparticles in combination that results in the $T_\varphi$ trend that we observe here.

Alternatively, if we assume that the charge noise in the qubit is limited to around $1\times10^{-3}$~\si{e\per\sqrt{\hertz}} for all qubits, which is generally assumed for transmons, then there is a significant additional effect that is limiting the dephasing times in the sample, likely from regions away from the Josephson junction.
This could be due to the TLSs present in the capacitor pads or the ground near the qubit.

\section{Fridge wiring diagram}\label{app:fridge}

\begin{figure*}
    \centering
    \includegraphics[width=0.8\linewidth]{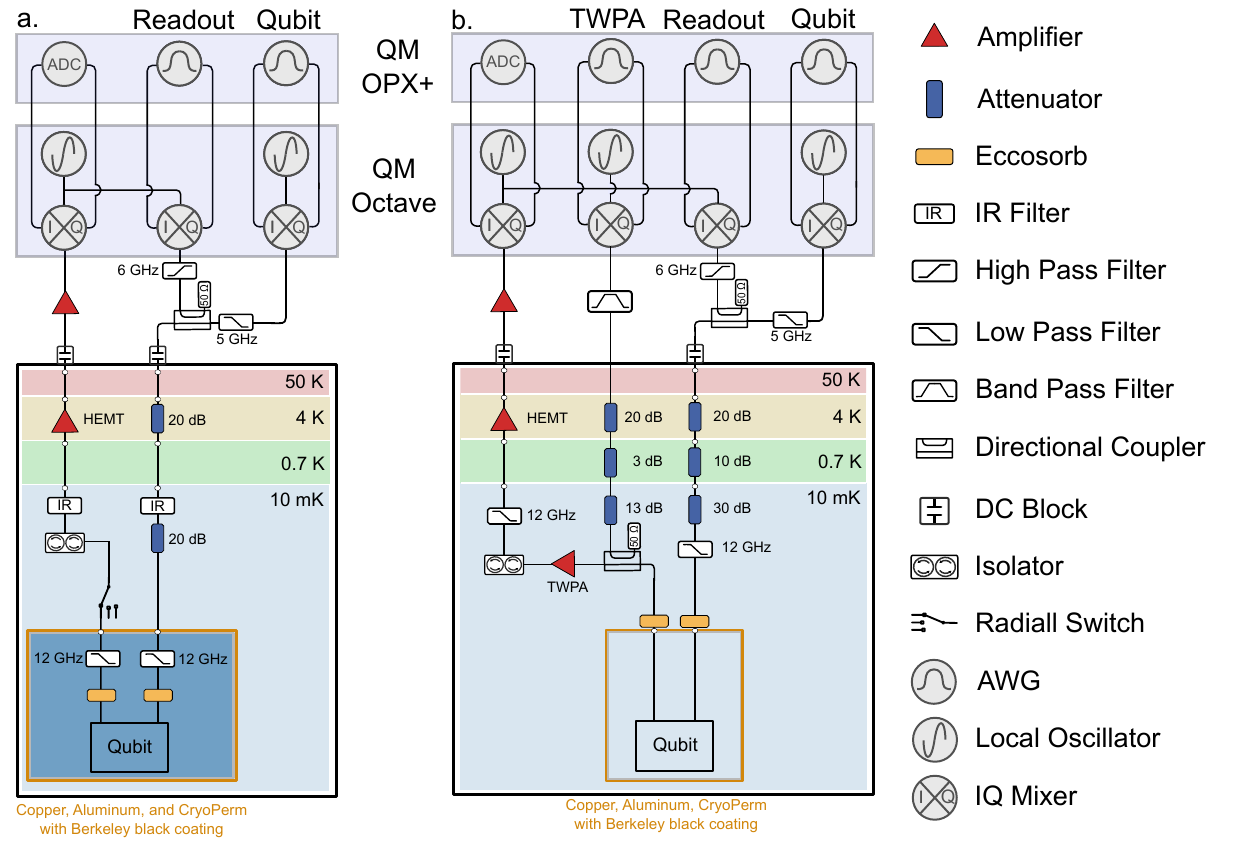}
    \caption{Fridge wiring diagrams for (a) cooldown 1 in fridge 1, and (b) cooldown 2 in fridge 2.
    }
    \label{fig:app_wiring}
\end{figure*}

The wiring diagram for the two Bluefors LD250 fridges used for qubit measurement are shown in Fig.~\ref{fig:app_wiring}.
Both fridges has a base temperature below \SI{10}{\milli\kelvin}.
Both measurements were conducted with the same Quantum Machines OPX+ and Octave.
The addition of room temperature filters and the TWPA line did not significantly affect the measurements in any meaningful way.
Any variation, although can be due to variations between fridges, can also be explained by the two week room temperature aging of the qubit.

\section{Further fluctuation data}\label{app:fluctuations}

Allan deviation and power spectral density (PSD) analysis were conducted on the same time trace data. 
Overlapping Allan deviation was extracted by using the AllanTools package (v2024.6)~\cite{wallin_aewallin_2026}, with the maximum integration time set to T\textsubscript{total}/4 to ensure statistical significance.
The Allan deviation is presented in units of~\si{\micro\second} to be compared to the $T_1$ being measured. 

PSD analysis was conducted using the Welch method on the time trace with the mean subtracted ($T_1-\mu_{T_1}$).
The spectral density calculation was done with 128 points in each segment and a periodic Hann window. 
1/f trend line is plotted by using the second datapoint of the PSD as the initialization, followed by a slope of $-1$.
The white noise line in both plots has the same amplitude, defined as $A_W =6\times10^3$~\si{\square\micro\second\per\hertz} in the PSD graph, and $\sqrt{A_W/\tau}$ in the Allan deviation plot. Note that neither of these are fitted in the analysis, and present for illustrative purposes as a reference towards understanding the trends observed within the plots.

\begin{figure*}
    \centering
    \includegraphics[width=\linewidth]{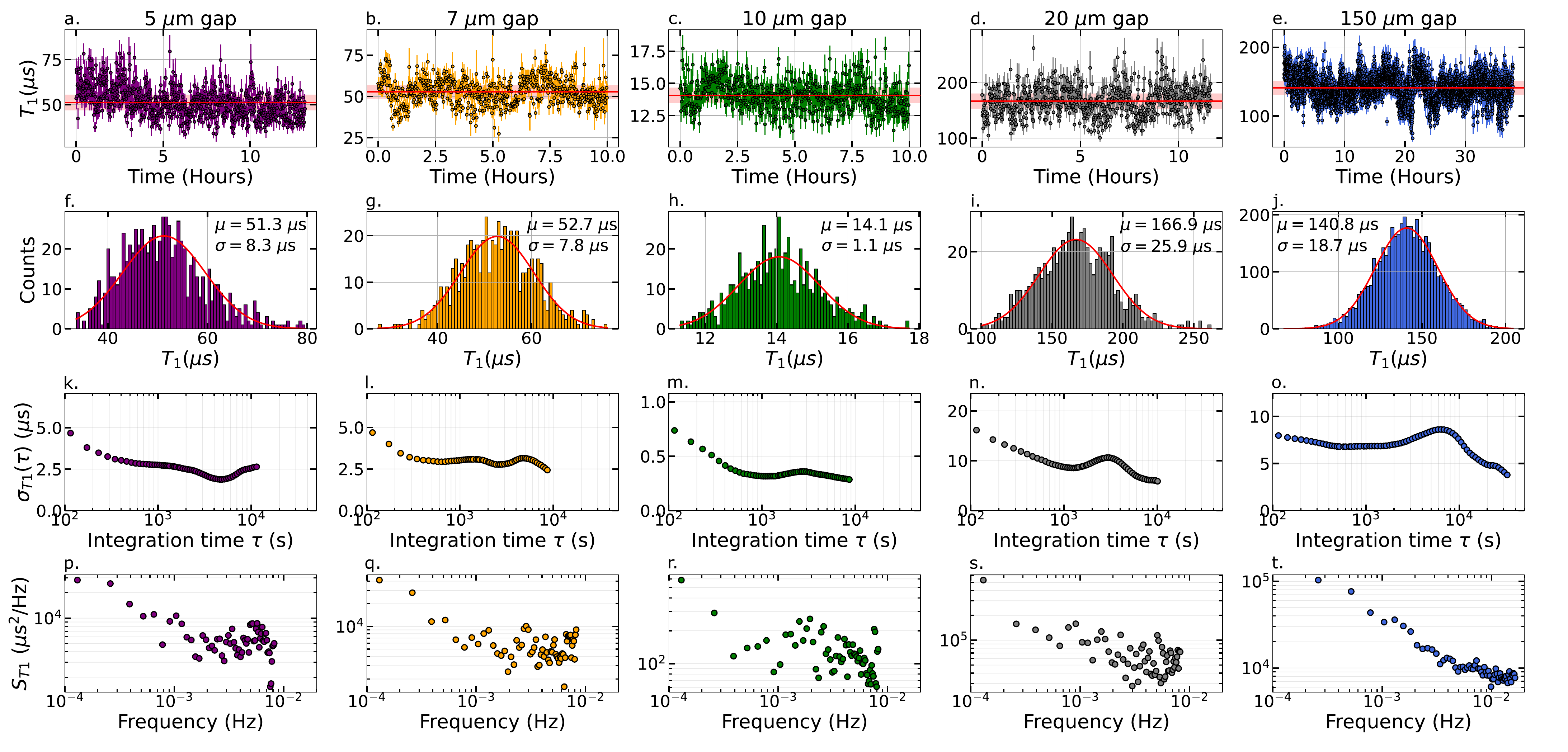}
    \caption{$T_1$ fluctuation analysis of all 5 qubits. 
    (a)-(e) $T_1$ time traces, with the mean and standard deviation marked in red.
    (f)-(j) Histogram of the $T_1$ time traces with an overlaid Gaussian distribution (red) with calculated mean $\mu$ and standard deviation $\sigma$.
    (k)-(o) Overlapping Allan deviation with a sliding window of $T_{\rm total}/4$.
    (p)-(t) Power spectral density calculated using the Welch method with 128 points per segment. 
    }
    \label{fig:app_T1fluctuations}
\end{figure*}

We conduct the overlapping Allan deviation and PSD  statistical analysis on the $T_1$ fluctuations of all 5 qubits in the second cooldown, hereby referred to as Q-5, Q-7, Q-10, Q-20, and Q-150, denoting their capacitor gap size in microns.
The complete dataset is shown in Fig.~\ref{fig:app_T1fluctuations}. 
Note that the time interval between individual measurements is \SI{30}{\second} for Q-150, while it is \SI{60}{\second} for all others.
Also note the difference in the total time of data collection of Q-150 compared to the rest.
Data from Q-150 therefore provides better statistics and lower noise compared to the other four.

Q-5, Q-7, and Q-20 have a reduced $T_1$ compared to the first cooldown, by roughly \SI{10}{\micro\second}, \SI{15}{\micro\second}, and \SI{17}{\micro\second} respectively. 
We see that all 5 time traces show switching events where the $T_1$ deviates, usually downwards (Fig.~\ref{fig:app_T1fluctuations}(a)-(e)).
Regardless, the $T_1$ fluctuations remain in a mostly Gaussian distribution (Fig.~\ref{fig:app_T1fluctuations}(f)-(j)), implying that the any TLS-related fluctuation does not significantly affect the overall shape of the Gaussian.
The overlapping Allan deviation shows notable differences between the 5 qubits (Fig.~\ref{fig:app_T1fluctuations}(k)-(o)).
All qubits have deviations that scale roughly with the $T_1$, showing similar ratios to the Q-150, indicating that the reduction in $T_1$ fluctuation is systematic rather than specific to Q-150.
The deviation of Q-20, which has the highest deviation of the qubits measured, is still lower than a comparable qubit~\cite{berritta_realtime_2026}, highlighting that the overall reduction in fluctuation of the trench-based qubit still holds.

Another clear difference within the Allan deviation plots is the appearance of more significant (or multiple) peaks within the curve, indicative of Lorentzian noise processes.
As established in~\cite{berritta_realtime_2026,burnett_decoherence_2019}, these Lorentzian components originate from telegraph switching due to strongly coupled TLS.
This is an indicator of near-resonant TLSs inducing changes in the coherence time on those respective timescales.
Q-150 does not contain any significant peak below $\sim10^4$~\si{\second}, which as mentioned in the main text, could be due to a near-resonant TLS or some external environmental noise. 

Finally, examining the PSD plots for all qubits (Fig.~\ref{fig:app_T1fluctuations}(p)-(t)), we see that they are all significantly noisier compared to Q-150 due to the 5-6$\times$ reduction in the number of datapoints.
They all seem to show 1/f-like behavior initially, followed by a saturation due to the white noise floor.
Any Lorentzian process would be difficult to identify within the PSD plot.
The frequency resolution is around \SI{200}{\micro\hertz} for the PSD plot, which already shows significant noise. It is possible that there could be TLS-associated noise at higher frequency resolution, which would also be hidden by the noise.

\begin{figure*}
    \centering
    \includegraphics[width=\linewidth]{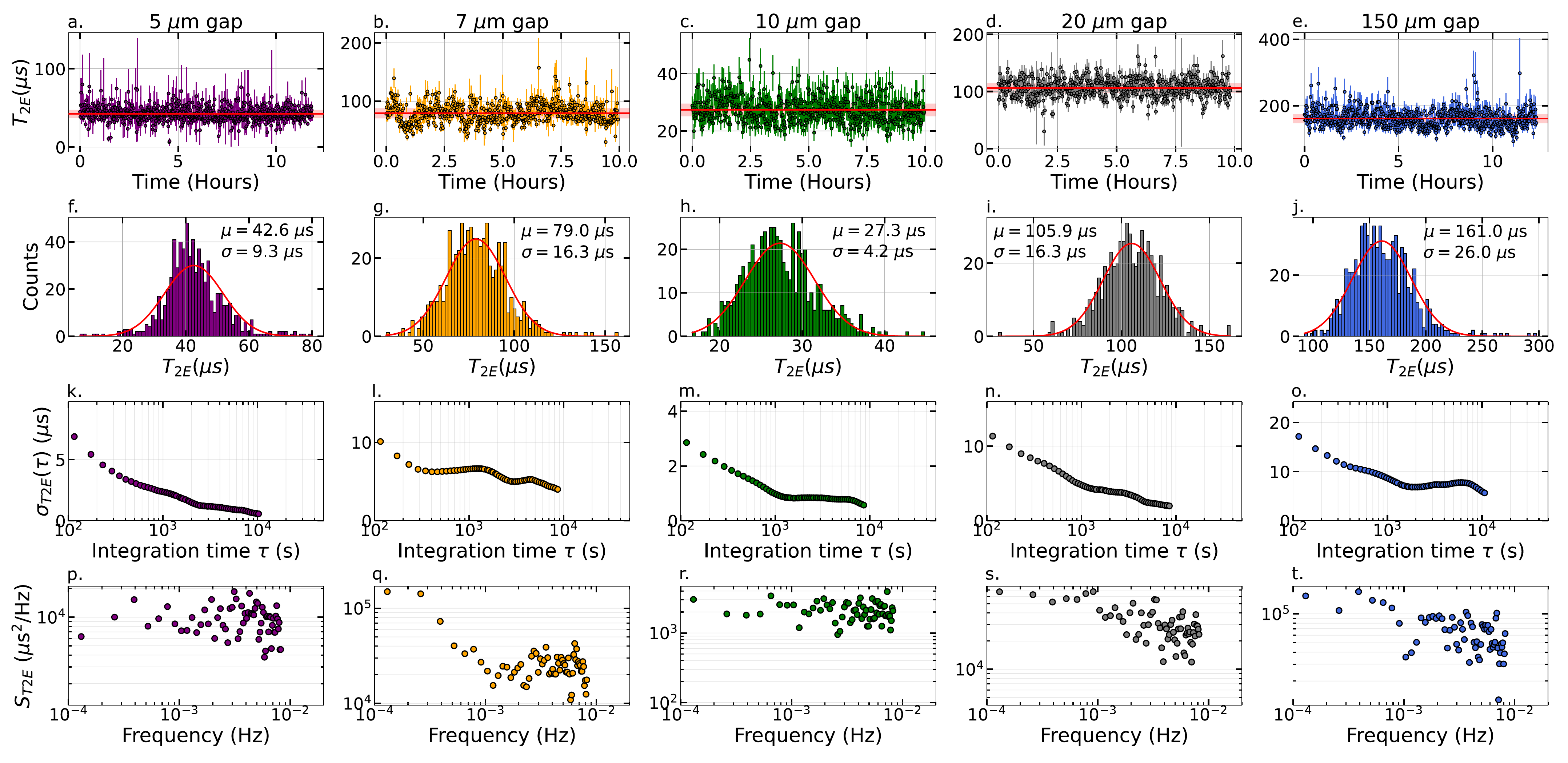}
    \caption{$T_{2E}$ fluctuation analysis of all 5 qubits. 
    (a)-(e) $T_{2E}$ time traces, with the mean and standard deviation marked in red.
    (f)-(j) Histogram of the  $T_{2E}$ time traces with an overlaid Gaussian distribution (red) with calculated mean $\mu$ and standard deviation $\sigma$.
    (k)-(o) Overlapping Allan deviation with a sliding window of $T_{\rm total}/4$.
    (p)-(t) Power spectral density calculated using the Welch method with 128 points per segment. 
    }
    \label{fig:app_T2Efluctuations}
\end{figure*}

We also conduct identical analyses on the $T_{2E}$ data from the 5 qubits, shown in Fig.~\ref{fig:app_T2Efluctuations}.
For Q-5, Q-7, Q-10, and Q-20, the $T_{2E}$ data was collected in an interleaved manner with the $T_1$ data in Fig.~\ref{fig:app_T1fluctuations}.
For Q-150, the $T_{2E}$ data was collected at a different time to the data in Fig.~\ref{fig:app_T1fluctuations}.

The data shows that the $T_{2E}$ has changed for Q-20 and Q-150 from the first cooldown by a notable amount.
The distributions have a larger number of outliers compared to the $T_1$ fluctuations, resulting in broader tails in all distributions.
Not all distributions are symmetric and Gaussian in this instance, with Q-5 being narrower (Fig.~\ref{fig:app_T2Efluctuations}(f)) and Q-10 having a skew towards higher values (Fig.~\ref{fig:app_T2Efluctuations}(i)).

It is notable that the Allan deviation for all 5 qubits is higher for the $T_{2E}$ fluctuations compared to the $T_1$ fluctuations. 
They still mimic the features seen in the $T_1$ plots, such as the two peaks in Q-7 or the peak in Q-150.
The PSD data is similarly noisy as in Fig.~\ref{fig:app_T1fluctuations}, and the 1/f-like behavior is not seen in Q-5 or Q-10, which seem to be at the white noise floor.
The others show some decay, most notably in Q-7, which matches the $T_1$ PSD curve quite closely.

\section{Robust coefficient of variance comparison}\label{app:RCV}

$T_1$ time traces are reported on more frequently, alongside standard statistical information from box plots such as medians and interquartile range (IQR).
This enables a comparison against other works using the robust coefficient of variation (RCV)~\cite{botta-dukat_quartile_2023}, defined as the ratio of the IQR (first-third quartile) to the median.
This is an incomplete metric for comparison, as these numbers depend significantly on the number of datapoints, sampling rate, and many other factors that affect such measurements.
Regardless, this can be used as a simple measure to have quantitative comparison.
For papers with published box plots, IQR and median were directly extracted from the plots~\cite{olszewski_kryptonsputtered_2026, gordon_environmental_2022,kono_mechanically_2024,colaozanuz_mitigating_2025,bland_millisecond_2025,klimov_fluctuations_2018}.
For papers that contain histograms or mean and standard deviation information only, the data was assumed to be Gaussian and symmetric, and the IQR was calculated as $\approx1.349\sigma$~\cite{biznarova_mitigation_2024,tuokkola_methods_2025,burnett_decoherence_2019}.
We note that this assumption is inaccurate, as the IQR may not be coupled to the $\sigma$ well due to the high possibility of skewed or non-Gaussianity of the $T_1$ distributions, in which case the RCV calculate will be higher than the true value.
Regardless, the works that do contain this information are nominally Gaussian, and we assume that the information is provided with some degree of representativeness of the overall distribution.
Therefore, the IQR and $\sigma$ should be correlated to provide a relatively representative RCV.

\begin{figure}
    \centering
    \includegraphics[width=\linewidth]{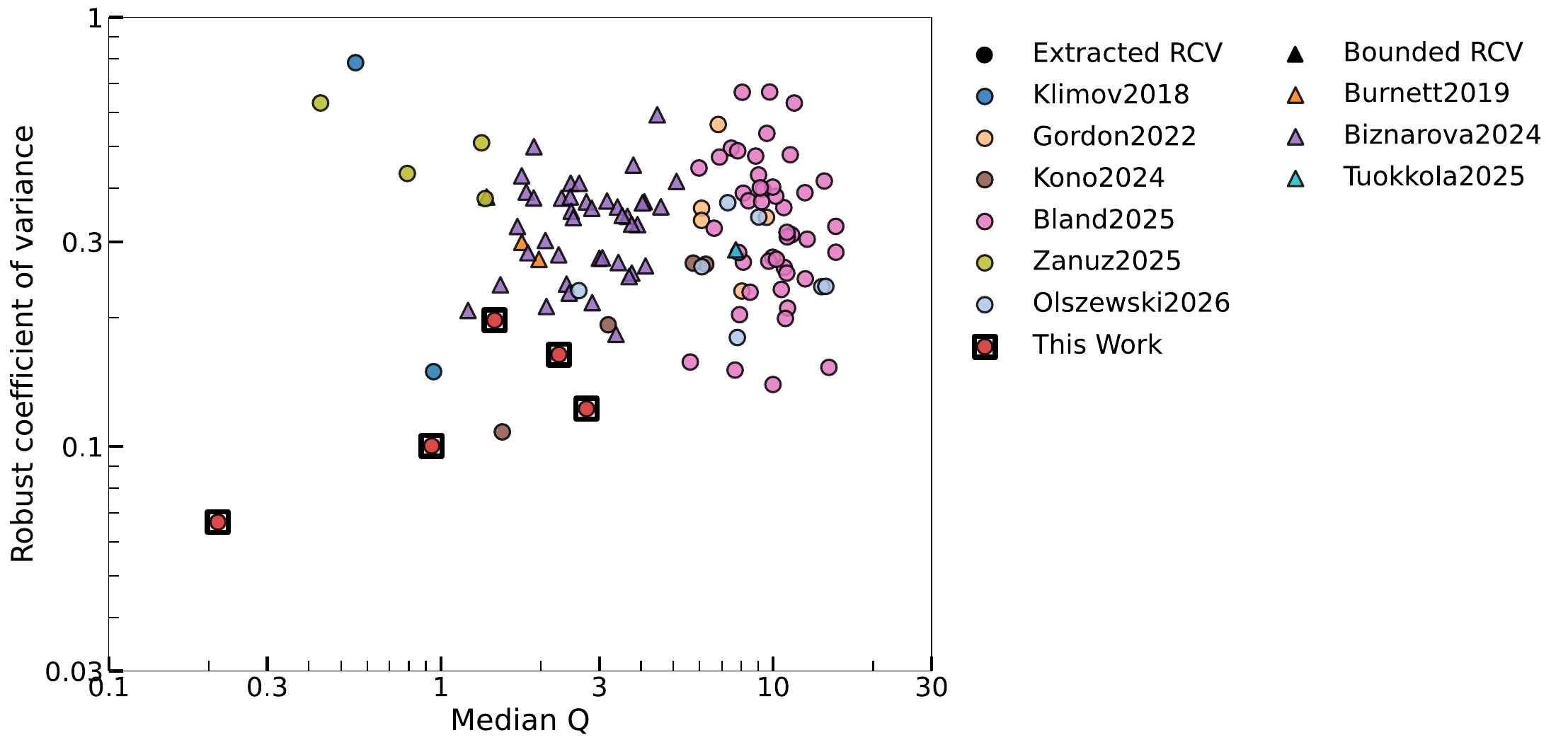}
    \caption{Robust coefficient of variance (RCV) as a function of quality factor Q for select recent works in comparison to the work done here~\cite{klimov_fluctuations_2018,burnett_decoherence_2019, gordon_environmental_2022,biznarova_mitigation_2024,kono_mechanically_2024, bland_millisecond_2025,colaozanuz_mitigating_2025,tuokkola_methods_2025,olszewski_kryptonsputtered_2026}. 
    RCV was either extracted directly from box plots in the respective works (Extracted RCV, circles) or calculated assuming a Gaussian distribution using the given mean and standard deviation, which provides a reasonable lower bound (Bounded RCV, triangles).
    RCV tends to be at or above 0.2 for most works.
    The highest RCV from the qubits measured in this work is 0.19.
    }
    \label{fig:app_comparison}
\end{figure}

Fig.~\ref{fig:app_comparison} plots the RCV as a function of quality factor Q for a number of works in the field in recent years~\cite{klimov_fluctuations_2018,burnett_decoherence_2019, gordon_environmental_2022, biznarova_mitigation_2024, kono_mechanically_2024, bland_millisecond_2025,colaozanuz_mitigating_2025,tuokkola_methods_2025,olszewski_kryptonsputtered_2026}. 
The RCV tends to have values above 0.2 for most works regardless of the median Q, although they contain outliers that fall below the 0.2 value.
The 5 qubits measured here show RCV consistently below 0.2 (highest being 0.19).
This suggests that the qubits made using the trench process results in lower fluctuations compared to recent works with comparable quality factors. 
Moving forward, standardization of time-fluctuation measurement and analysis methods will help to improve comparability.

\clearpage

\bibliography{references_TB}

@misc{abdisatarov_demonstrating_2025,
  title = {Demonstrating Magnetic Field Robustness and Reducing Temporal {{T1}} Noise in Transmon Qubits through Magnetic Field Engineering},
  author = {Abdisatarov, Bektur and Roy, Tanay and Bafia, Daniel and Pilipenko, Roman and Dubiel, Matthew Julian and van Zanten, David and Zhu, Shaojiang and Bal, Mustafa and Eremeev, Grigory and {Elsayed-Ali}, Hani and Murty, Akshay and Romanenko, Alexander and Grassellino, Anna},
  year = 2025,
  month = jun,
  number = {arXiv:2506.02187},
  eprint = {2506.02187},
  primaryclass = {quant-ph},
  publisher = {arXiv},
  doi = {10.48550/arXiv.2506.02187},
  urldate = {2026-03-28},
  archiveprefix = {arXiv}
}

@article{ambegaokar_tunneling_1963,
  title = {Tunneling {{Between Superconductors}}},
  author = {Ambegaokar, Vinay and Baratoff, Alexis},
  year = 1963,
  month = jun,
  journal = {Physical Review Letters},
  volume = {10},
  number = {11},
  pages = {486--489},
  issn = {0031-9007},
  doi = {10.1103/PhysRevLett.10.486},
  urldate = {2024-08-08},
  copyright = {http://link.aps.org/licenses/aps-default-license},
  langid = {english}
}

@article{anferov_improved_2024,
  title = {Improved Coherence in Optically Defined Niobium Trilayer-Junction Qubits},
  author = {Anferov, Alexander and Lee, Kan-Heng and Zhao, Fang and Simon, Jonathan and Schuster, David I.},
  year = 2024,
  month = feb,
  journal = {Physical Review Applied},
  volume = {21},
  number = {2},
  pages = {024047},
  issn = {2331-7019},
  doi = {10.1103/PhysRevApplied.21.024047},
  urldate = {2024-08-05},
  langid = {english}
}

@article{anferov_superconducting_2024,
  title = {Superconducting {{Qubits}} above 20 {{GHz Operating}} over 200 {{mK}}},
  author = {Anferov, Alexander and Harvey, Shannon P. and Wan, Fanghui and Simon, Jonathan and Schuster, David I.},
  year = 2024,
  month = sep,
  journal = {PRX Quantum},
  volume = {5},
  number = {3},
  pages = {030347},
  issn = {2691-3399},
  doi = {10.1103/PRXQuantum.5.030347},
  urldate = {2026-03-29},
  langid = {english}
}

@misc{antonenko_effect_2025,
  title = {Effect of Quasiparticles on the Parameters of a Gap-Engineered Transmon},
  author = {Antonenko, Daniil S. and Kurilovich, Pavel D. and {Matute-Ca{\~n}adas}, Francisco J. and Glazman, Leonid I.},
  year = 2025,
  publisher = {arXiv},
  doi = {10.48550/ARXIV.2507.23169},
  urldate = {2026-05-08},
  copyright = {arXiv.org perpetual, non-exclusive license}
}

@article{bal_systematic_2024,
  title = {Systematic Improvements in Transmon Qubit Coherence Enabled by Niobium Surface Encapsulation},
  author = {Bal, Mustafa and Murthy, Akshay A. and Zhu, Shaojiang and Crisa, Francesco and You, Xinyuan and Huang, Ziwen and Roy, Tanay and Lee, Jaeyel and Zanten, David Van and Pilipenko, Roman and Nekrashevich, Ivan and Lunin, Andrei and Bafia, Daniel and Krasnikova, Yulia and Kopas, Cameron J. and Lachman, Ella O. and Miller, Duncan and Mutus, Josh Y. and Reagor, Matthew J. and Cansizoglu, Hilal and Marshall, Jayss and Pappas, David P. and Vu, Kim and Yadavalli, Kameshwar and Oh, Jin-Su and Zhou, Lin and Kramer, Matthew J. and Lecocq, Florent and Goronzy, Dominic P. and {Torres-Castanedo}, Carlos G. and Pritchard, P. Graham and Dravid, Vinayak P. and Rondinelli, James M. and Bedzyk, Michael J. and Hersam, Mark C. and Zasadzinski, John and Koch, Jens and Sauls, James A. and Romanenko, Alexander and Grassellino, Anna},
  year = 2024,
  month = apr,
  journal = {npj Quantum Information},
  volume = {10},
  number = {1},
  pages = {43},
  issn = {2056-6387},
  doi = {10.1038/s41534-024-00840-x},
  urldate = {2024-08-05},
  langid = {english}
}

@misc{banerjee_dataset_2026,
  title = {Dataset for ''{{Resist-free}} Shadow Deposition Using Silicon Trenches for {{Josephson}} Junctions in Superconducting Qubits"},
  author = {Banerjee, Tathagata and Fatemi, Valla},
  year = 2026,
  publisher = {Zenodo},
  doi = {10.5281/zenodo.19303492}
}

@article{banerjee_fabrication_2026,
  title = {Fabrication Effects on Niobium Oxidation and Surface Contamination in Niobium-Metal Bilayers Using x-Ray Photoelectron Spectroscopy},
  author = {Banerjee, Tathagata and Olszewski, Maciej W. and Fatemi, Valla},
  year = 2026,
  month = apr,
  journal = {AIP Advances},
  volume = {16},
  number = {4},
  pages = {045028},
  issn = {2158-3226},
  doi = {10.1063/5.0325433},
  urldate = {2026-04-30},
  langid = {english}
}

@article{berritta_realtime_2026,
  title = {Real-{{Time Adaptive Tracking}} of {{Fluctuating Relaxation Rates}} in {{Superconducting Qubits}}},
  author = {Berritta, Fabrizio and Benestad, Jacob and Krzywda, Jan A. and Krause, Oswin and Marciniak, Malthe A. and Kr{\o}jer, Svend and Warren, Christopher W. and Hogedal, Emil and Nylander, Andreas and Ahmad, Irshad and Osman, Amr and Bizn{\'a}rov{\'a}, Janka and Rommel, Marcus and Roudsari, Anita Fadavi and Bylander, Jonas and Tancredi, Giovanna and Danon, Jeroen and Hastrup, Jacob and Kuemmeth, Ferdinand and Kjaergaard, Morten},
  year = 2026,
  month = feb,
  journal = {Physical Review X},
  volume = {16},
  number = {1},
  pages = {011025},
  issn = {2160-3308},
  doi = {10.1103/gk1b-stl3},
  urldate = {2026-03-26},
  langid = {english}
}

@article{biznarova_mitigation_2024,
  title = {Mitigation of Interfacial Dielectric Loss in Aluminum-on-Silicon Superconducting Qubits},
  author = {Bizn{\'a}rov{\'a}, Janka and Osman, Amr and Rehnman, Emil and Chayanun, Lert and Kri{\v z}an, Christian and Malmberg, Per and Rommel, Marcus and Warren, Christopher and Delsing, Per and Yurgens, August and Bylander, Jonas and Fadavi Roudsari, Anita},
  year = 2024,
  month = aug,
  journal = {npj Quantum Information},
  volume = {10},
  number = {1},
  pages = {78},
  issn = {2056-6387},
  doi = {10.1038/s41534-024-00868-z},
  urldate = {2026-02-10},
  langid = {english}
}

@article{bland_millisecond_2025,
  title = {Millisecond Lifetimes and Coherence Times in {{2D}} Transmon Qubits},
  author = {Bland, Matthew P. and Bahrami, Faranak and Martinez, Jeronimo G. C. and Prestegaard, Paal H. and Smitham, Basil M. and Joshi, Atharv and Hedrick, Elizabeth and Kumar, Shashwat and Yang, Ambrose and {Pakpour-Tabrizi}, Alexander C. and Jindal, Apoorv and Chang, Ray D. and Cheng, Guangming and Yao, Nan and Cava, Robert J. and De Leon, Nathalie P. and Houck, Andrew A.},
  year = 2025,
  month = nov,
  journal = {Nature},
  volume = {647},
  number = {8089},
  pages = {343--348},
  issn = {0028-0836, 1476-4687},
  doi = {10.1038/s41586-025-09687-4},
  urldate = {2026-01-08},
  langid = {english}
}

@article{botta-dukat_quartile_2023,
  title = {Quartile Coefficient of Variation Is More Robust than {{CV}} for Traits Calculated as a Ratio},
  author = {{Botta-Duk{\'a}t}, Zolt{\'a}n},
  year = 2023,
  month = mar,
  journal = {Scientific Reports},
  volume = {13},
  number = {1},
  pages = {4671},
  issn = {2045-2322},
  doi = {10.1038/s41598-023-31711-8},
  urldate = {2026-03-28},
  langid = {english}
}

@article{burnett_decoherence_2019,
  title = {Decoherence Benchmarking of Superconducting Qubits},
  author = {Burnett, Jonathan J. and Bengtsson, Andreas and Scigliuzzo, Marco and Niepce, David and Kudra, Marina and Delsing, Per and Bylander, Jonas},
  year = 2019,
  month = jun,
  journal = {npj Quantum Information},
  volume = {5},
  number = {1},
  pages = {54},
  issn = {2056-6387},
  doi = {10.1038/s41534-019-0168-5},
  urldate = {2026-02-19},
  langid = {english}
}

@article{choi_low_2025,
  title = {Low Barrier {{ZrO}} {\emph{x}} -Based {{Josephson}} Junctions},
  author = {Choi, Jaehong and Olszewski, Maciej and Zhang, Luojia and Baraissov, Zhaslan and Banerjee, Tathagata and Aggarwal, Kushagra and Chaudhari, Sarvesh and Arias, Tom{\'a}s A. and Muller, David A. and Fatemi, Valla and Fuchs, Gregory D.},
  year = 2025,
  month = nov,
  journal = {APL Materials},
  volume = {13},
  number = {11},
  pages = {111103},
  issn = {2166-532X},
  doi = {10.1063/5.0296881},
  urldate = {2025-11-07},
  langid = {english}
}

@article{christensen_anomalous_2019,
  title = {Anomalous Charge Noise in Superconducting Qubits},
  author = {Christensen, B. G. and Wilen, C. D. and Opremcak, A. and Nelson, J. and Schlenker, F. and Zimonick, C. H. and Faoro, L. and Ioffe, L. B. and Rosen, Y. J. and DuBois, J. L. and Plourde, B. L. T. and McDermott, R.},
  year = 2019,
  month = oct,
  journal = {Physical Review B},
  volume = {100},
  number = {14},
  pages = {140503},
  issn = {2469-9950, 2469-9969},
  doi = {10.1103/PhysRevB.100.140503},
  urldate = {2026-04-07},
  langid = {english}
}

@article{colaozanuz_mitigating_2025,
  title = {Mitigating Losses of Superconducting Qubits Strongly Coupled to Defect Modes},
  author = {Colao Zanuz, Dante and Ficheux, Quentin and Michaud, Laurent and Orekhov, Alexei and Hanke, Kilian and Flasby, Alexander and Bahrami Panah, Mohsen and Norris, Graham J. and Kerschbaum, Michael and Remm, Ants and Swiadek, Fran{\c c}ois and Hellings, Christoph and Laz{\u a}r, Stefania and Scarato, Colin and Lacroix, Nathan and Krinner, Sebastian and Eichler, Christopher and Wallraff, Andreas and Besse, Jean-Claude},
  year = 2025,
  month = apr,
  journal = {Physical Review Applied},
  volume = {23},
  number = {4},
  pages = {044054},
  publisher = {American Physical Society (APS)},
  issn = {2331-7019},
  doi = {10.1103/physrevapplied.23.044054},
  urldate = {2025-05-01},
  copyright = {https://link.aps.org/licenses/aps-default-license},
  langid = {english}
}

@misc{dane_performance_2025,
  title = {Performance {{Stabilization}} of {{High-Coherence Superconducting Qubits}}},
  author = {Dane, Andrew and Balakrishnan, Karthik and Wacaser, Brent and Hung, Li-Wen and Mamin, H. J. and Rugar, Daniel and Shelby, Robert M. and Murray, Conal and Rodbell, Kenneth and Sleight, Jeffrey},
  year = 2025,
  month = mar,
  number = {arXiv:2503.12514},
  eprint = {2503.12514},
  primaryclass = {quant-ph},
  publisher = {arXiv},
  doi = {10.48550/arXiv.2503.12514},
  urldate = {2025-03-26},
  archiveprefix = {arXiv}
}

@misc{ezratty_there_2023,
  title = {Is There a {{Moore}}'s Law for Quantum Computing?},
  author = {Ezratty, Olivier},
  year = 2023,
  month = mar,
  number = {arXiv:2303.15547},
  eprint = {2303.15547},
  primaryclass = {quant-ph},
  publisher = {arXiv},
  doi = {10.48550/arXiv.2303.15547},
  urldate = {2026-03-28},
  archiveprefix = {arXiv}
}

@misc{funni_sdfunni_2025,
  title = {Sdfunni/{{SingleOrigin}}: {{SingleOrigin}} v3.0b2},
  shorttitle = {Sdfunni/{{SingleOrigin}}},
  author = {Funni, Stephen D and Charles Evans},
  year = 2025,
  month = dec,
  doi = {10.5281/ZENODO.18100473},
  urldate = {2026-03-27},
  copyright = {GNU General Public License v3.0 only},
  howpublished = {Zenodo}
}

@article{gambetta_investigating_2017,
  title = {Investigating {{Surface Loss Effects}} in {{Superconducting Transmon Qubits}}},
  author = {Gambetta, Jay M. and Murray, Conal E. and Fung, Y.-K.-K. and McClure, Douglas T. and Dial, Oliver and Shanks, William and Sleight, Jeffrey W. and Steffen, Matthias},
  year = 2017,
  month = jan,
  journal = {IEEE Transactions on Applied Superconductivity},
  volume = {27},
  number = {1},
  pages = {1--5},
  issn = {1051-8223, 1558-2515},
  doi = {10.1109/TASC.2016.2629670},
  urldate = {2026-03-28},
  copyright = {https://ieeexplore.ieee.org/Xplorehelp/downloads/license-information/OAPA.html}
}

@misc{gingras_improving_2025,
  title = {Improving {{Transmon Qubit Performance}} with {{Fluorine-based Surface Treatments}}},
  author = {Gingras, Michael A. and Niedzielski, Bethany M. and Grossklaus, Kevin A. and Miller, Duncan and Contipelli, Felipe and Azar, Kate and Burkhart, Luke D. and Calusine, Gregory and Davis, Daniel and Pi{\~n}ero, Ren{\'e}e DePencier and Gertler, Jeffrey M. and Hazard, Thomas M. and Hirjibehedin, Cyrus F. and Kim, David K. and Knecht, Jeffrey M. and Melville, Alexander J. and O'Connell, Christopher and Rood, Robert A. and Sabbah, Ali and Stickler, Hannah and Yoder, Jonilyn L. and Oliver, William D. and Schwartz, Mollie E. and Serniak, Kyle},
  year = 2025,
  month = jul,
  number = {arXiv:2507.08089},
  eprint = {2507.08089},
  primaryclass = {quant-ph},
  publisher = {arXiv},
  doi = {10.48550/arXiv.2507.08089},
  urldate = {2026-01-08},
  archiveprefix = {arXiv}
}

@article{googlequantumaiandcollaborators_quantum_2025,
  title = {Quantum Error Correction below the Surface Code Threshold},
  author = {{Google Quantum AI and Collaborators}},
  year = 2025,
  month = feb,
  journal = {Nature},
  volume = {638},
  number = {8052},
  pages = {920--926},
  doi = {10.1038/s41586-024-08449-y}
}

@article{gordon_environmental_2022,
  title = {Environmental Radiation Impact on Lifetimes and Quasiparticle Tunneling Rates of Fixed-Frequency Transmon Qubits},
  author = {Gordon, R. T. and Murray, C. E. and Kurter, C. and Sandberg, M. and Hall, S. A. and Balakrishnan, K. and Shelby, R. and Wacaser, B. and Stabile, A. A. and Sleight, J. W. and Brink, M. and Rothwell, M. B. and Rodbell, K. P. and Dial, O. and Steffen, M.},
  year = 2022,
  month = feb,
  journal = {Applied Physics Letters},
  volume = {120},
  number = {7},
  pages = {074002},
  issn = {0003-6951, 1077-3118},
  doi = {10.1063/5.0078785},
  urldate = {2026-03-28},
  langid = {english}
}

@article{hanna_onchip_2026,
  title = {On-Chip Stencil Lithography for Superconducting Qubits},
  author = {Hanna, Roudy and Ihssen, S{\"o}ren and Geisert, Simon and Kocak, Umut and Arfini, Matteo and Hertel, Albert and Smart, Thomas J. and Schleenvoigt, Michael and Schmitt, Tobias and Domnick, Joscha and Underwood, Kaycee and Jalil, Abdur Rehman and Bae, Jin Hee and Bennemann, Benjamin and F{\'e}chant, Mathieu and Field, Mitchell and Spiecker, Martin and Zapata, Nicolas and Dickel, Christian and Berenschot, Erwin and Tas, Niels and Steele, Gary A. and Gr{\"u}tzmacher, Detlev and Pop, Ioan M. and Sch{\"u}ffelgen, Peter},
  year = 2026,
  month = jun,
  journal = {Applied Physics Reviews},
  volume = {13},
  number = {2},
  pages = {021403},
  issn = {1931-9401},
  doi = {10.1063/5.0307532},
  urldate = {2026-04-30},
  langid = {english}
}

@article{houck_controlling_2008,
  title = {Controlling the {{Spontaneous Emission}} of a {{Superconducting Transmon Qubit}}},
  author = {Houck, A. A. and Schreier, J. A. and Johnson, B. R. and Chow, J. M. and Koch, Jens and Gambetta, J. M. and Schuster, D. I. and Frunzio, L. and Devoret, M. H. and Girvin, S. M. and Schoelkopf, R. J.},
  year = 2008,
  month = aug,
  journal = {Physical Review Letters},
  volume = {101},
  number = {8},
  pages = {080502},
  issn = {0031-9007, 1079-7114},
  doi = {10.1103/PhysRevLett.101.080502},
  urldate = {2026-04-03},
  copyright = {http://link.aps.org/licenses/aps-default-license},
  langid = {english}
}

@article{ithier_decoherence_2005,
  title = {Decoherence in a Superconducting Quantum Bit Circuit},
  author = {Ithier, G. and Collin, E. and Joyez, P. and Meeson, P. J. and Vion, D. and Esteve, D. and Chiarello, F. and Shnirman, A. and Makhlin, Y. and Schriefl, J. and Sch{\"o}n, G.},
  year = 2005,
  month = oct,
  journal = {Physical Review B},
  volume = {72},
  number = {13},
  pages = {134519},
  issn = {1098-0121, 1550-235X},
  doi = {10.1103/PhysRevB.72.134519},
  urldate = {2026-04-21},
  copyright = {http://link.aps.org/licenses/aps-default-license},
  langid = {english}
}

@article{klimov_fluctuations_2018,
  title = {Fluctuations of {{Energy-Relaxation Times}} in {{Superconducting Qubits}}},
  author = {Klimov, P. V. and Kelly, J. and Chen, Z. and Neeley, M. and Megrant, A. and Burkett, B. and Barends, R. and Arya, K. and Chiaro, B. and Chen, Yu and Dunsworth, A. and Fowler, A. and Foxen, B. and Gidney, C. and Giustina, M. and Graff, R. and Huang, T. and Jeffrey, E. and Lucero, Erik and Mutus, J. Y. and Naaman, O. and Neill, C. and Quintana, C. and Roushan, P. and Sank, Daniel and Vainsencher, A. and Wenner, J. and White, T. C. and Boixo, S. and Babbush, R. and Smelyanskiy, V. N. and Neven, H. and Martinis, John M.},
  year = 2018,
  month = aug,
  journal = {Physical Review Letters},
  volume = {121},
  number = {9},
  pages = {090502},
  issn = {0031-9007, 1079-7114},
  doi = {10.1103/PhysRevLett.121.090502},
  urldate = {2026-02-20},
  langid = {english}
}

@article{koch_chargeinsensitive_2007,
  title = {Charge-Insensitive Qubit Design Derived from the {{Cooper}} Pair Box},
  author = {Koch, Jens and Yu, Terri M. and Gambetta, Jay and Houck, A. A. and Schuster, D. I. and Majer, J. and Blais, Alexandre and Devoret, M. H. and Girvin, S. M. and Schoelkopf, R. J.},
  year = 2007,
  month = oct,
  journal = {Physical Review A},
  volume = {76},
  number = {4},
  pages = {042319},
  issn = {1050-2947, 1094-1622},
  doi = {10.1103/PhysRevA.76.042319},
  urldate = {2025-02-06},
  copyright = {http://link.aps.org/licenses/aps-default-license},
  langid = {english}
}

@article{kono_mechanically_2024,
  title = {Mechanically Induced Correlated Errors on Superconducting Qubits with Relaxation Times Exceeding 0.4 Ms},
  author = {Kono, Shingo and Pan, Jiahe and Chegnizadeh, Mahdi and Wang, Xuxin and Youssefi, Amir and Scigliuzzo, Marco and Kippenberg, Tobias J.},
  year = 2024,
  month = may,
  journal = {Nature Communications},
  volume = {15},
  number = {1},
  pages = {3950},
  issn = {2041-1723},
  doi = {10.1038/s41467-024-48230-3},
  urldate = {2026-02-10},
  langid = {english}
}

@misc{kopas_simple_2022,
  title = {Simple Coplanar Waveguide Resonator Mask Targeting Metal-Substrate Interface},
  author = {Kopas, Cameron J. and Lachman, Ella and McRae, Corey Rae H. and Mohan, Yuvraj and Mutus, Josh Y. and Nersisyan, Ani and Poudel, Amrit},
  year = 2022,
  month = apr,
  number = {arXiv:2204.07202},
  eprint = {2204.07202},
  primaryclass = {quant-ph},
  publisher = {arXiv},
  doi = {10.48550/arXiv.2204.07202},
  urldate = {2025-06-03},
  archiveprefix = {arXiv}
}

@article{krantz_quantum_2019,
  title = {A Quantum Engineer's Guide to Superconducting Qubits},
  author = {Krantz, P. and Kjaergaard, M. and Yan, F. and Orlando, T. P. and Gustavsson, S. and Oliver, W. D.},
  year = 2019,
  month = jun,
  journal = {Applied Physics Reviews},
  volume = {6},
  number = {2},
  pages = {021318},
  issn = {1931-9401},
  doi = {10.1063/1.5089550},
  urldate = {2025-06-10},
  langid = {english}
}

@misc{krizan_electrical_2026,
  title = {Electrical Post-Fabrication Tuning of Aluminum {{Josephson}} Junctions at Room Temperature},
  author = {Kri{\v z}an, Christian and Toselli, Maurizio and Ahmad, Irshad and Khaksaran, Hadi and Rommel, Marcus and Trnjanin, Nermin and Bizn{\'a}rov{\'a}, Janka and Dahiya, Mamta and Hogedal, Emil and Jakobsson, Halld{\'o}r and Nylander, Andreas and Bylander, Jonas and Delsing, Per and Tancredi, Giovanna},
  year = 2026,
  month = feb,
  number = {arXiv:2602.20002},
  eprint = {2602.20002},
  primaryclass = {quant-ph},
  publisher = {arXiv},
  doi = {10.48550/arXiv.2602.20002},
  urldate = {2026-04-04},
  archiveprefix = {arXiv}
}

@article{kurilovich_correlated_2026,
  title = {Correlated {{Phase Error Bursts}} in a {{Gap-Engineered Superconducting Qubit Array}}},
  author = {Kurilovich, Vladislav D. and Roberts, Gabrielle and Martin, Leigh S. and McEwen, Matt and Eickbusch, Alec and Faoro, Lara and Ioffe, Lev B. and Atalaya, Juan and Bilmes, Alexander and Kreikebaum, John Mark and Bengtsson, Andreas and Klimov, Paul and Neeley, Matthew and Mruczkiewicz, Wojciech and Miao, Kevin and Aleiner, Igor L. and Kelly, Julian and Chen, Yu and Satzinger, Kevin and Opremcak, Alex},
  year = 2026,
  month = may,
  journal = {Physical Review X},
  volume = {16},
  number = {2},
  pages = {021025},
  issn = {2160-3308},
  doi = {10.1103/1bl4-b2f7},
  urldate = {2026-05-11},
  langid = {english}
}

@article{lazar_enabling_2025,
  title = {Enabling Electron-Energy-Loss Spectroscopy at Very High Energy Losses: {{An}} Opportunity to Obtain x-Ray Absorption Spectroscopy--like Information Using an Electron Microscope},
  shorttitle = {Enabling Electron-Energy-Loss Spectroscopy at Very High Energy Losses},
  author = {Lazar, Sorin and Tiemeijer, Peter and Schnohr, Claudia S. and Meledina, Maria and Patzig, Christian and H{\"o}che, Thomas and Longo, Paolo and Freitag, Bert},
  year = 2025,
  month = may,
  journal = {Physical Review Applied},
  volume = {23},
  number = {5},
  pages = {054095},
  issn = {2331-7019},
  doi = {10.1103/PhysRevApplied.23.054095},
  urldate = {2026-03-27},
  langid = {english}
}

@article{matityahu_qubit_2024,
  title = {Qubit Dephasing by Spectrally Diffusing Quantum Two-Level Systems},
  author = {Matityahu, Shlomi and Shnirman, Alexander and Schechter, Moshe},
  year = 2024,
  month = apr,
  journal = {Physical Review Applied},
  volume = {21},
  number = {4},
  pages = {044055},
  issn = {2331-7019},
  doi = {10.1103/PhysRevApplied.21.044055},
  urldate = {2026-05-13},
  langid = {english}
}

@article{murray_analytical_2020,
  title = {Analytical {{Modeling}} of {{Participation Reduction}} in {{Superconducting Coplanar Resonator}} and {{Qubit Designs Through Substrate Trenching}}},
  author = {Murray, Conal E.},
  year = 2020,
  month = aug,
  journal = {IEEE Transactions on Microwave Theory and Techniques},
  volume = {68},
  number = {8},
  pages = {3263--3270},
  issn = {0018-9480, 1557-9670},
  doi = {10.1109/TMTT.2020.2995894},
  urldate = {2026-04-03},
  copyright = {https://ieeexplore.ieee.org/Xplorehelp/downloads/license-information/IEEE.html}
}

@article{murray_material_2021,
  title = {Material Matters in Superconducting Qubits},
  author = {Murray, Conal E.},
  year = 2021,
  month = oct,
  journal = {Materials Science and Engineering: R: Reports},
  volume = {146},
  pages = {100646},
  issn = {0927796X},
  doi = {10.1016/j.mser.2021.100646},
  urldate = {2024-06-08},
  langid = {english}
}

@article{nakamura_coherent_1999,
  title = {Coherent Control of Macroscopic Quantum States in a Single-{{Cooper-pair}} Box},
  author = {Nakamura, Y. and Pashkin, {\relax Yu}. A. and Tsai, J. S.},
  year = 1999,
  month = apr,
  journal = {Nature},
  volume = {398},
  number = {6730},
  pages = {786--788},
  issn = {0028-0836, 1476-4687},
  doi = {10.1038/19718},
  urldate = {2026-03-29},
  copyright = {http://www.springer.com/tdm},
  langid = {english}
}

@misc{olszewski_kryptonsputtered_2026,
  title = {Krypton-Sputtered Tantalum Films for Scalable High-Performance Quantum Devices},
  author = {Olszewski, Maciej W. and Kong, Lingda and Reinhardt, Simon and Tong, Daniel and Du, Xinyi and Gianluca, Gabriele Di and Lu, Haoran and Roy, Saswata and Zhang, Luojia and Biedron, Aleksandra B. and Muller, David A. and Fatemi, Valla},
  year = 2026,
  month = jan,
  number = {arXiv:2601.20091},
  eprint = {2601.20091},
  primaryclass = {quant-ph},
  publisher = {arXiv},
  doi = {10.48550/arXiv.2601.20091},
  urldate = {2026-02-12},
  archiveprefix = {arXiv}
}

@misc{olszewski_lowloss_2025,
  title = {Low-Loss {{Nb}} on {{Si}} Superconducting Resonators from a Dual-Use Spintronics Deposition Chamber and with Acid-Free Post-Processing},
  author = {Olszewski, Maciej W. and Paustian, Jadrien T. and Banerjee, Tathagata and Lu, Haoran and Ramirez, Jorge L. and Nguyen, Nhi and Okubo, Kiichi and Pant, Rohit and Biedron, Aleksandra B. and Ralph, Daniel C. and Richardson, Christopher J. K. and Fuchs, Gregory D. and McRae, Corey Rae H. and Pechenezhskiy, Ivan V. and Plourde, B. L. T. and Fatemi, Valla},
  year = 2025,
  month = mar,
  number = {arXiv:2503.13285},
  eprint = {2503.13285},
  primaryclass = {quant-ph},
  publisher = {arXiv},
  doi = {10.48550/arXiv.2503.13285},
  urldate = {2025-03-25},
  archiveprefix = {arXiv}
}

@article{place_new_2021,
  title = {New Material Platform for Superconducting Transmon Qubits with Coherence Times Exceeding 0.3 Milliseconds},
  author = {Place, Alexander P. M. and Rodgers, Lila V. H. and Mundada, Pranav and Smitham, Basil M. and Fitzpatrick, Mattias and Leng, Zhaoqi and Premkumar, Anjali and Bryon, Jacob and Vrajitoarea, Andrei and Sussman, Sara and Cheng, Guangming and Madhavan, Trisha and Babla, Harshvardhan K. and Le, Xuan Hoang and Gang, Youqi and J{\"a}ck, Berthold and Gyenis, Andr{\'a}s and Yao, Nan and Cava, Robert J. and De Leon, Nathalie P. and Houck, Andrew A.},
  year = 2021,
  month = mar,
  journal = {Nature Communications},
  volume = {12},
  number = {1},
  pages = {1779},
  issn = {2041-1723},
  doi = {10.1038/s41467-021-22030-5},
  urldate = {2024-02-22},
  langid = {english}
}

@article{pop_fabrication_2012,
  title = {Fabrication of Stable and Reproducible Submicron Tunnel Junctions},
  author = {Pop, I. M. and Fournier, T. and Crozes, T. and Lecocq, F. and Matei, I. and Pannetier, B. and Buisson, O. and Guichard, W.},
  year = 2012,
  month = jan,
  journal = {Journal of Vacuum Science \& Technology B, Nanotechnology and Microelectronics: Materials, Processing, Measurement, and Phenomena},
  volume = {30},
  number = {1},
  pages = {010607},
  issn = {2166-2746, 2166-2754},
  doi = {10.1116/1.3673790},
  urldate = {2024-08-08},
  langid = {english}
}

@techreport{riley_handbook_2008,
  title = {Handbook of Frequency Stability Analysis},
  author = {Riley, W J},
  year = 2008,
  edition = {0},
  number = {NIST SP 1065},
  pages = {NIST SP 1065},
  address = {Gaithersburg, MD},
  institution = {{National Institute of Standards and Technology}},
  doi = {10.6028/NIST.SP.1065},
  urldate = {2026-03-31},
  langid = {english}
}

@article{sears_photon_2012,
  title = {Photon Shot Noise Dephasing in the Strong-Dispersive Limit of Circuit {{QED}}},
  author = {Sears, A. P. and Petrenko, A. and Catelani, G. and Sun, L. and Paik, Hanhee and Kirchmair, G. and Frunzio, L. and Glazman, L. I. and Girvin, S. M. and Schoelkopf, R. J.},
  year = 2012,
  month = nov,
  journal = {Physical Review B},
  volume = {86},
  number = {18},
  pages = {180504},
  issn = {1098-0121, 1550-235X},
  doi = {10.1103/PhysRevB.86.180504},
  urldate = {2026-04-06},
  copyright = {http://link.aps.org/licenses/aps-default-license},
  langid = {english}
}

@article{sivak_realtime_2023,
  title = {Real-Time Quantum Error Correction beyond Break-Even},
  author = {Sivak, V. V. and Eickbusch, A. and Royer, B. and Singh, S. and Tsioutsios, I. and Ganjam, S. and Miano, A. and Brock, B. L. and Ding, A. Z. and Frunzio, L. and Girvin, S. M. and Schoelkopf, R. J. and Devoret, M. H.},
  year = 2023,
  month = apr,
  journal = {Nature},
  volume = {616},
  number = {7955},
  pages = {50--55},
  issn = {0028-0836, 1476-4687},
  doi = {10.1038/s41586-023-05782-6},
  urldate = {2026-03-29},
  langid = {english}
}

@article{tsioutsios_freestanding_2020,
  title = {Free-Standing Silicon Shadow Masks for Transmon Qubit Fabrication},
  author = {Tsioutsios, I. and Serniak, K. and Diamond, S. and Sivak, V. V. and Wang, Z. and Shankar, S. and Frunzio, L. and Schoelkopf, R. J. and Devoret, M. H.},
  year = 2020,
  month = jun,
  journal = {AIP Advances},
  volume = {10},
  number = {6},
  pages = {065120},
  issn = {2158-3226},
  doi = {10.1063/1.5138953},
  urldate = {2026-03-26},
  langid = {english}
}

@article{tuokkola_methods_2025,
  title = {Methods to Achieve Near-Millisecond Energy Relaxation and Dephasing Times for a Superconducting Transmon Qubit},
  author = {Tuokkola, Mikko and Sunada, Yoshiki and Kivij{\"a}rvi, Heidi and Albanese, Jonatan and Gr{\"o}nberg, Leif and Kaikkonen, Jukka-Pekka and Vesterinen, Visa and Govenius, Joonas and M{\"o}tt{\"o}nen, Mikko},
  year = 2025,
  month = jul,
  journal = {Nature Communications},
  volume = {16},
  number = {1},
  pages = {5421},
  issn = {2041-1723},
  doi = {10.1038/s41467-025-61126-0},
  urldate = {2026-01-08},
  langid = {english}
}

@article{verjauw_path_2022,
  title = {Path toward Manufacturable Superconducting Qubits with Relaxation Times Exceeding 0.1 Ms},
  author = {Verjauw, J. and Acharya, R. and Van Damme, J. and Ivanov, {\relax Ts}. and Lozano, D. Perez and Mohiyaddin, F. A. and Wan, D. and Jussot, J. and Vadiraj, A. M. and Mongillo, M. and Heyns, M. and Radu, I. and Govoreanu, B. and Poto{\v c}nik, A.},
  year = 2022,
  month = aug,
  journal = {npj Quantum Information},
  volume = {8},
  number = {1},
  pages = {93},
  issn = {2056-6387},
  doi = {10.1038/s41534-022-00600-9},
  urldate = {2026-03-29},
  langid = {english}
}

@misc{wallin_aewallin_2026,
  title = {Aewallin/Allantools},
  author = {Wallin, Anders},
  year = 2026,
  month = mar,
  urldate = {2026-03-31},
  copyright = {LGPL-3.0}
}

@phdthesis{wang_fabrication_2015,
  title = {Fabrication Stability of {{Josephson}} Junctions for Superconducting Qubits},
  author = {Wang, Lujun},
  year = 2015,
  langid = {english},
  school = {Technische Universitat Munchen}
}

@article{wang_surface_2015,
  title = {Surface Participation and Dielectric Loss in Superconducting Qubits},
  author = {Wang, C. and Axline, C. and Gao, Y. Y. and Brecht, T. and Chu, Y. and Frunzio, L. and Devoret, M. H. and Schoelkopf, R. J.},
  year = 2015,
  month = oct,
  journal = {Applied Physics Letters},
  volume = {107},
  number = {16},
  pages = {162601},
  issn = {0003-6951, 1077-3118},
  doi = {10.1063/1.4934486},
  urldate = {2026-03-28},
  langid = {english}
}

@article{welch_use_1967,
  title = {The Use of Fast {{Fourier}} Transform for the Estimation of Power Spectra: {{A}} Method Based on Time Averaging over Short, Modified Periodograms},
  shorttitle = {The Use of Fast {{Fourier}} Transform for the Estimation of Power Spectra},
  author = {Welch, P.},
  year = 1967,
  month = jun,
  journal = {IEEE Transactions on Audio and Electroacoustics},
  volume = {15},
  number = {2},
  pages = {70--73},
  issn = {0018-9278},
  doi = {10.1109/TAU.1967.1161901},
  urldate = {2026-03-31},
  copyright = {https://ieeexplore.ieee.org/Xplorehelp/downloads/license-information/IEEE.html},
  langid = {english}
}

@misc{wolff_structural_2026,
  title = {Structural Control of Two-Level Defect Density Revealed by High-Throughput Correlative Measurements of {{Josephson}} Junctions},
  author = {Wolff, Oliver F. and Mantry, Harshvardhan and Raja, Rahim and Peng, Wei-Hsiang and Singirikonda, Kaushik and Lee, Seungkyun and Sudhaman, Shishir and Goncalves, Rafael and Huang, Pinshane Y. and Kou, Angela and Pfaff, Wolfgang},
  year = 2026,
  month = feb,
  number = {arXiv:2602.11469},
  eprint = {2602.11469},
  primaryclass = {quant-ph},
  publisher = {arXiv},
  doi = {10.48550/arXiv.2602.11469},
  urldate = {2026-04-07},
  archiveprefix = {arXiv}
}

@article{yan_flux_2016,
  title = {The Flux Qubit Revisited to Enhance Coherence and Reproducibility},
  author = {Yan, Fei and Gustavsson, Simon and Kamal, Archana and Birenbaum, Jeffrey and Sears, Adam P and Hover, David and Gudmundsen, Ted J. and Rosenberg, Danna and Samach, Gabriel and Weber, S and Yoder, Jonilyn L. and Orlando, Terry P. and Clarke, John and Kerman, Andrew J. and Oliver, William D.},
  year = 2016,
  month = nov,
  journal = {Nature Communications},
  volume = {7},
  number = {1},
  pages = {12964},
  issn = {2041-1723},
  doi = {10.1038/ncomms12964},
  urldate = {2026-04-06},
  langid = {english}
}

@article{zeng_direct_2015,
  title = {Direct Observation of the Thickness Distribution of Ultra Thin {{AlO}}{\textsubscript{ {\emph{x}} }} Barriers in {{Al}}/{{AlO}}{\textsubscript{ {\emph{x}} }} /{{Al Josephson}} Junctions},
  author = {Zeng, L J and Nik, S and Greibe, T and Krantz, P and Wilson, C M and Delsing, P and Olsson, E},
  year = 2015,
  month = oct,
  journal = {Journal of Physics D: Applied Physics},
  volume = {48},
  number = {39},
  pages = {395308},
  issn = {0022-3727, 1361-6463},
  doi = {10.1088/0022-3727/48/39/395308},
  urldate = {2026-04-07}
}

\end{document}